\documentclass[11pt,preprint]{aastex}

\usepackage{lscape}
\usepackage{epsfig}

\newcommand{\ebv}{E$_{B-V}$}
\newcommand{\caii}{Ca~{\small II}}
\newcommand{\sfh}{$\mathcal{SFH}$}
\newcommand{\eby}{E$_{b-y}$}
\newcommand{\evby}{E$_{m1}$}
\newcommand{\vbby}{$m_1$}
\newcommand{\vhb}{V$-$V$_{\mathrm{HB}}$}
\newcommand{\teff}{T$_{\mathrm{eff}}$}
\newcommand{\dmh}{$\Delta \left[\frac{M}{H}\right]$}

\shorttitle{LMC Metallicity Distribution Function}
\shortauthors{Cole, Smecker-Hane and Gallagher}

\begin{document}

\title{The Metallicity Distribution Function of Red Giants
in the LMC\altaffilmark{1,3}}

\author{Andrew A. Cole\altaffilmark{2}}
\affil{Department of Astronomy, 532A Lederle Grad Research Tower,
University of Massachusetts, Amherst, MA 01003;
{\it cole@condor.astro.umass.edu}}

\author{Tammy A. Smecker-Hane\altaffilmark{2}}
\affil{Department of Physics and Astronomy, University of California
at Irvine, 4129 Frederick Reines Hall, Irvine, CA 92697--4575;
{\it smecker@carina.ps.uci.edu}}

\and

\author{John S. Gallagher, III}
\affil{Department of Astronomy, University of Wisconsin--Madison,
5534 Sterling Hall, 475 North Charter Street, Madison, WI 53706--1582;
{\it jsg@astro.wisc.edu}}

\altaffiltext{1}{Based on observations obtained at
Cerro Tololo Inter-American Observatory, a division of the
National Optical Astronomy Observatories, which are operated
by the Association of Universities for Research in
Astronomy, Inc. under cooperative agreement with the
National Science Foundation.}
\altaffiltext{2}{Visiting Astronomer, Cerro Tololo Inter-American Observatory.}
\altaffiltext{3}{To appear in the October, 2000 issue of The Astronomical 
Journal.}

\begin{abstract}
We report new metallicity determinations for 39 red giants
in a 220 square arcminute region, 1$\fdg$8 southwest of the bar of the Large
Magellanic Cloud.  These abundance measurements are based on 
spectroscopy of the \caii\ infrared triplet.
We have carefully considered the effects of abundance ratios, the
physics of \caii\ line formation, the variation of red clump magnitude,
and the contamination by foreground stars in our abundance analyses.
The metallicity distribution
function (MDF) shows a strong peak at [Fe/H] = $-$0.57 $\pm$0.04; a tail 
to abundances at least as low as [Fe/H] $\approx$ $-$1.6 brings the
average abundance down to [Fe/H] = $-$0.64 $\pm$0.02.  Half the red
giants in our field fall within the range $-$0.83 $\leq$ [Fe/H] 
$\leq$ $-$0.41.  The MDF appears 
to be truncated at [Fe/H] $\approx$ $-$0.25; the exact value of the
maximum abundance is subject to $\sim$0.1 dex uncertainty in the
calibration of the
\caii\ IR triplet for young, metal-rich stars.
We find a striking contrast in the shape of the MDF below [Fe/H] $\leq$
$-$1 between our inner disk field and
the distant outer field studied by Olszewski (1993): red giants deficient
by more than a factor of ten in heavy elements relative to the Sun are
extremely scarce in the inner disk of the LMC.  Our field star sample
does not reproduce the full MDF of the LMC star clusters, but seems
similar to that of the intermediate-age (1--3 Gyr) clusters.
We have also obtained abundance estimates using Str\"{o}mgren 
photometry for $\approx$10$^3$ red giants in the 
same field.  Photometry is the only practical way to measure abundances
for the large numbers of stars necessary to lift age-metallicity
degeneracy from our high-precision color-magnitude diagrams.
The Str\"{o}mgren measurements, which are sensitive to a 
combination of cyanogen and iron lines, correlate well with the 
\caii\ measurements, but a metallicity-dependent offset is found.
The offset may be due either to variations in the elemental abundance
ratios due to galactic chemical evolution, or to a metal-dependent
mixing mechanism in RGB stars.  An empirical relation between 
photometric and spectroscopic abundance estimates is derived.  This
will allow photometric abundance measurements to be placed
on a consistent metallicity scale with spectroscopic metallicities,
for very large numbers of stars.
\end{abstract}


\section{Introduction}

The Large Magellanic Cloud (LMC) is a critical system for the
understanding of how galaxies evolve.  The past history of a galaxy
is best read from a detailed examination of its stellar content;
in the LMC, individual stars may be resolved to a limiting mass of 
$\approx$0.6 M$_{\sun}$ with the {\it Hubble Space Telescope} \citep{gal96}.
The prospect
therefore exists that we will soon be able to uniquely and completely 
determine its star-formation history (\sfh) based on
deep color-magnitude diagram analyses \citep{sme99}.  The 
LMC is a keystone for studies of intermediate-age (1--10 Gyr)
stellar populations, because as a galaxy it is dominated by stars that
formed in a massive event some 2--5 Gyr ago \citep{but77,har84,str84}.
The development of instrumentation and techniques to study the chemical
evolution of the LMC exponentially increases the amount of information
available to constrain its \sfh , 
by mitigating the age-metallicity degeneracy
of color-magnitude diagrams (e.g., Hodge 1989, Da Costa 1991,
Olszewski et al. 1996).

The great profusion of bright star clusters in the LMC has naturally 
attracted attention for several decades, and for many years it was
thought that the galaxy could be understood purely through the age
and abundance distribution of these objects \citep{baa63}.  However,
as early as 1964, Tifft noted in a review that truly ancient star clusters
seemed rare in the LMC, while the bulk of the stars down to V $\approx$
18 were apparently some few billion years old.  During the same time 
period, Gascoigne (1964) began to map the color distribution of star
clusters, establishing the first evidence for what would become 
known as the LMC cluster age gap \citep{dac91,wes97}. 

The unusual age distribution of the LMC star clusters
has been emphasized by van den Bergh (1998).  Of the
approximately 4500 clusters in the LMC, fewer than 10
fall into the age range 4--12 Gyr \citep{dac91,sar98}.
While the field star-formation rate
apparently increased by a factor of $\approx$2--3 some
2--4 Gyr ago (e.g., Holtzman et al. 1999 and references therein),
the rate of cluster formation has increased by nearly
two orders of magnitude during the same time period.
A substantial population of truly ancient clusters
is present \citep{sun92}, proving that
the lack of 4--12 Gyr-old clusters is not entirely due to the
operation of some cluster destruction mechanism during
this period.  The deficiency in moderately old, rich star clusters
reflects the true history of cluster formation and cannot wholly
be the artifact of incompleteness in surveys, or
systematic errors in age estimation techniques \citep{dac99}.
Da Costa (1991) and Olszewski et al. (1991) called attention
to the fact that the cluster age-gap coresponds precisely
with an abundance gap as well:  the $\gtrsim$12 Gyr-old
clusters have [Fe/H] $\approx$ $-$1.8, while the 1--3 Gyr-old
clusters have [Fe/H] $\approx$ $-$0.5.

Because of the cluster age/abundance gap, the clusters 
tell us nothing about the metallicity evolution of the LMC
from 3--12 Gyr ago.  The age-metallicity relation during
these ``dark ages'' of cluster production remain ill-constrained.
The first determination of the 
field star age-metallicity relation, based on $\alpha$-element
abundances in planetary nebulae \citep{dop97}, showed evidence
for almost no metallicity evolution prior to $\approx$2 Gyr ago,
when an abrupt doubling of the mean abundance occurred.  However,
just 5 planetaries older than $\approx$3 Gyr were included in 
the Dopita et al. (1997) sample; still, this remains the best guide to
the field star age-metallicity relation to date.

Owing to this uncertainty, a
gap in our knowledge of the intermediate-age \sfh\ of the LMC
persists (see Gallagher et al. 1999).
The age/abundance gap presents an apparent
paradox \citep{ols96}: how could the LMC have
been enriched in metals by a factor of $\approx$20 during
a time period in which it formed few to no stars?  The paradox
is resolved if the field star-formation rate is decoupled
from the cluster formation rate; this chemical evolution
constraint has spurred the investigation of the field stellar
populations of the LMC \citep{gal99,edo99}.  If the results
of Dopita et al. (1997) persist to larger stellar samples, indicating
that the field star chemical enrichment closely follows that of
the clusters, then simple evolutionary models for the LMC will 
be sorely challenged to explain the abundance patterns.

The state of knowledge of the LMC's intermediate-age field populations
has been summarized by Westerlund (1997) and Olszewski (1999); the 
application of high-quality, WFPC2 data to the problem has been
discussed by Gallagher et al. (1999) and Olsen (1999).  
While much recent progress has been
made, the basic picture hasn't changed much since
Hodge (1989) wrote his influential review on the stellar 
populations of the Local Group.  Notably: the LMC field stars
appear to be mostly of intermediate (2--4 Gyr) age, but the ratio
of older to younger stars remains uncertain by a factor of a few; and
although there is a clear trend of increasing metallicity with decreasing
age, there is a large scatter about the mean relation for a given age.
This last point was recently re-emphasized by Holtzman et al. (1999), who 
explored some of the systematic effects of the dispersion in the 
age-metallicity relation on determinations of the \sfh.
It is still somewhat uncertain as to how much of the perceived
dispersion in the age-metallicity relation is real, and how much
is attributable to systematic errors in observational methods. 

We have undertaken a long-term program to derive the \sfh\ of 
the field stars in the LMC disk and bar.  The 
primary dataset for this project is a suite of deep, high-quality
WFPC2 color-magnitude diagrams with over 80,000 stars per field
\citep{sme99}.  To analyze these CMDs fully will require independent
knowledge of the metallicity distribution function (MDF) of 
each field, and so we have
undertaken a program of spectroscopy and intermediate-band photometry
to this purpose.  Because our target fields span a range of environments
within the LMC (bar, disk, and thick disk), our abundance data
may also address the question of radial gradients in metallicity
or the mixture of stellar populations.
The gradient of {\it mean} abundance is generally thought to
be small in the LMC
\citep{ols93,wes97}, however, the {\it shape} of the MDF for clusters
is observed to vary with radius \citep{bic99}; we will explore
the magnitude of this effect among the field stars.

We are interested in the MDF of LMC field stars in order to guide
our interpretation of deep WFPC2 color-magnitude diagrams, as a
means to determine the complete \sfh\ of the LMC.   As is well known,
CMD studies can be plagued by age-metallicity degeneracy, even for
star clusters, i.e., systems with a single epoch of star-formation at a
single constant metallicity.  In the compound stellar populations of a
galaxy like the LMC, this degeneracy can make the unique characterization of
SFR(t) and MDF(t) impossible.

Our eventual goal is to determine the star-formation rate of
the LMC, with a time resolution dt/t $\approx$ 10\% for its entire $\sim$14 Gyr
history, to an accuracy of $\pm$20\% per age bin.
This effort requires high-quality
photometric data for some 10$^4$ stars in the area of the main-sequence
turnoff and subgiant branch alone.  However, without independent
abundance information of comparable detail, the utility of even the
best CMDs is compromised.  In a CMD containing a total of
$\sim$10$^5$ stars, the only practical way to obtain abundance estimates
for even one percent of the stars is through the use of specially
designed photometric systems (e.g., Washington, DDO, Str\"{o}mgren, etc.).
Although a factor of 3--4 in precision, relative to spectroscopic
measurements, must be sacrificed for
individual stars, the large ensemble of measurements permits the
determination of the metallicity distribution function much more readily.
Spectroscopic and photometric observations complement each other in
this program, and we consider both approaches to our determination of
the MDF.
The observed MDF can then be taken as a convolution of the star-formation
rate, initial mass function, and age-metallicity relation Z(t)
over the star-forming lifetime of the LMC.  These quantities are
easily tabulated when trial solutions for the \sfh\ of the LMC based on
synthetic CMDs are obtained.
Requiring consistency between the observed and model MDF is
a powerful way to determine a unique \sfh\ solution.

The purpose of this paper is twofold: to present new estimates 
of metallicity distribution function (MDF) of red giants in a 
field of the inner disk of the LMC, and to evaluate the 
applicability of intermediate-band photometry in the Str\"{o}mgren
system to the problem of abundance determinations for large,
heterogeneous samples of evolved stars.  We present infrared
spectra of 39 red giants, at the wavelength of the \caii\ IR triplet, 
in \S \ref{casec}.  The metallicity calibration of this technique is
well-determined (\S \ref{r97sec}), although some uncertainty is
introduced by the possible loss of sensitivity at high metallicity,
possible age-dependance of the calibration, and uncertainty in the
patterns of variation of [Ca/Fe] with [Fe/H] among galaxies
with disparate star-formation histories (\S \ref{corsec}).
Although these spectra are of sufficient precision and accuracy
for our purposes, to obtain similar data for samples of $\sim$10$^3$
stars would consume an inconveniently large amount of telescope
time.  Therefore intermediate-band photometry is an attractive
alternative.  We describe our intermediate-band
photometric observations, reductions, and analysis in \S \ref{phsec}.
We derive a small mean value for the interstellar reddening, \ebv, 
along the line of sight to our disk field, and discuss 
discrepant results from the literature concerning the response
of the Str\"{o}mgren color indices to interstellar reddening, in 
\S \ref{ircsec}.  The determination of a accurate abundances for 
[Fe/H] $\leq$ $-$1 depends on a recent calibration
of the Str\"{o}mgren color-metallicity relation published by 
Hilker (2000) (\S \ref{cmrsec}).   We consider the effects of
differential reddening and contamination by Galactic foreground
dwarfs in \S \ref{drsec} and \S \ref{fgsec}.

We compare the spectroscopic and photometric MDFs in \S \ref{dsec},
and find good agreement between the two, reinforcing confidence in
the accuracy of both methods.  However, there is a metallicity-dependent
offset between the abundances derived for 34 stars in common to the
two samples.  The source of the discrepancy is likely to be related
to the sensitivity of the Str\"{o}mgren $m_1$ index to CN abundance
(Anthony-Twarog et al.\ 1995).

Regardless of calibration uncertainties,
we find a striking difference in the form of the MDF in our inner disk 
field, compared to that of the outer disk field studied by Olszewski (1993)
(\S \ref{varsec}).  Our results are summarized and their application
to ongoing studies of the \sfh\ of the LMC is discussed in \S \ref{sumsec}.

\section{\caii\ Spectroscopy \label{casec}}

It is important to obtain spectroscopic abundance measurements 
that are well-calibrated to a 
to a widely used abundance scale.  This will facilitate direct 
comparison of the MDFs between different fields and between field
stars and clusters.  The red giants of the LMC lie between 
16.5 $\leq$ I $\leq$ 20, and are thus unsuitable targets for high-resolution
spectroscopy from 4-meter class telescopes.  However, if sufficiently
strong lines are observed, abundances can be readily estimated using
moderate ($\sim$1 \AA) resolution spectrographs on 4-meter class telescopes
(e.g., Olszewski et al. 1991).

The peak of the red giant spectral energy distribution
lies in the near-infrared.  However, this region of the spectrum is strongly
contaminated by telluric absorption features and photospheric line
blends due to metal oxides that complicate the measurement
of individual absorption lines of interest.  The triplet of 
lines of singly-ionized calcium at $\lambda \lambda$ 8498, 8542, 8662 \AA\
are relatively easy to measure, because they lie neatly between 
atmospheric absorption bands of O$_2$ and H$_2$O and are the
strongest spectral features in the near-infrared \citep{jon84}.  The 
\caii\ triplet also falls near strong night sky emission lines
and thus sky contamination is a strong source of noise for faint stars.

Studies of the \caii\ triplet in metal-rich stars \citep{jon84}
and in the integrated light of populations dominated by relatively
young and/or metal-rich stars \citep{bic87} demonstrated that the
equivalent width of \caii\ ($W$) is strongly dependent on stellar luminosity
class (surface gravity), but only weakly dependent on metallicity.
However, for old populations more metal-poor than [Fe/H] $\approx$ 
$-$0.5, Bica \& Alloin (1987) found that metallicity is the controlling
parameter for $W$.  This was confirmed
by Armandroff \& Zinn (1988), who reported that the strength of the \caii\ 
feature in the integrated spectra of globular clusters varied in
lockstep with cluster metallicities.  

Since the early nineties, the \caii\ triplet has become the
spectral feature of choice for abundance measurements of
metal-poor systems, e.g. globular clusters 
\citep{arm91,dac95,rut97a}, dwarf spheroidal galaxies
\citep{leh92,sun93,dac95,sme99a}, and star clusters in the 
Magellanic Clouds \citep{ols91,sun92,dac98}.  The technique
is strictly calibrated only for old, metal-poor systems, but it
is roughly applicable to composite populations like the dSphs and
the Magellanic Clouds because its sensitivity to age is likely to
be small.  The most
comprehensive catalog of \caii\ measurements for globular 
clusters to date is given by Rutledge et al. (1997a), and the calibration
to the [Fe/H] scales of Zinn \& West (1984) and Carretta \& Gratton (1997)
is performed by Rutledge et al. (1997b).

All empirical and theoretical evidence indicates that $W$
is a stable and sensitive abundance indicator in old stellar systems; the
dependence on surface gravity can be corrected for by a simple
linear relation between $W$ and M$_V$, parameterized by the
difference in magnitude between the target star and the
level of the horizontal branch \citep{arm91,sun93,rut97b}.
However, the dependence of $W$ on surface gravity becomes highly
nonlinear for metallicities above [Ca/H] = $-$0.3 \citep{dia89};
this effect is magnified for intermediate-age stars \citep{jor92}.
Because many of the the stars in our sample are expected to be
much younger than the globular clusters
as well as more metal-rich \citep[this paper]{col99c}, we will
address the issues of corrections to the globular cluster-defined
metallicity calibration of $W$ in \S \ref{smsec}.

Olszewski et al. (1991) used the \caii\
triplet to obtain metallicities in 70 LMC star clusters,
deriving abundances accurate to $\pm$0.2 dex.  A much
larger effort to derive abundances for several hundred LMC field
stars near NGC 2257 (in the far northwestern part of the galaxy)
has also been undertaken \citep{ols93}, although only
preliminary results have been published to date (Olszewski 1993;
Suntzeff 1999, private communication).
A great advantage of our program
to use the \caii\ triplet is that we will be able to directly
compare our abundances to those of Olszewski et al. (1991) and
Olszewski (1993), testing for differences between clusters and field
stars, and for spatial variations in the abundance distribution.

\subsection{Observations, Reductions, \& Extractions \label{oresec}}

We obtained near-infrared spectra for 52 red stars in 
the ``Disk 1'' field (Smecker-Hane et al.\ 1999, also 
see Table \ref{fields}),
during three nights of bright time from 28--30 
December, 1998, using the Blanco 4-meter telescope
on Cerro Tololo.  This field lies in the inner disk
of the LMC, 1$\fdg$8 south of the optical center of
the bar.
Two-thirds of the first night, and two hours of
the third night were lost to equipment failure.  Useful science
exposures were obtained through the entire night of 29 December,
but poor atmospheric transparency limited the number of targets observed.  

We used the slit spectrograph at the Ritchey-Chr\'{e}tien focus with
the Blue Air Schmidt camera and Loral 3K CCD, yielding a spatial
scale of 0$\farcs$50/pixel.  The amplifiers were set to yield a gain of
1.9 e$^-$/ADU, with a readout noise of 7.5 e$^-$.  Only the central
700$\times$150 pixels of the CCD were read out, for higher data rate.
We observed all stars with the slit
decker fully open, permitting measurement of stars along the total
slit length of $\approx$5$\arcmin$.  With a slit width of 1$\farcs$5,
we achieved a spectral resolution of 1.1 \AA/pixel.  We used grating
G420, blazed at 8000 \AA\ with 600 lines/mm, in first order; a long-pass
filter with $\lambda _c$ $\approx$6000 \AA\ was used to block the
second order light. 
Our spectra covered the range 6700--9970 \AA.
Wavelength calibration was achieved using observations of 
He-Ne-Ar emission line standards; comparison lamp exposures were taken
between target exposures in order to ensure a stable wavelength solution
through the night.   

Targets were chosen based on their location in the CMD and on
favorable spatial positioning; in order to maximize observing
efficiency, the slit was rotated for each exposure to allow the
simultaneous measurement of 2--3 stars.  Exposure times for target
stars were 10--20 minutes; the relatively long integrations were
necessary due to clouds on night 2.  Each observing 
setup was designed to yield spectra for two LMC giants, identified
from the color-magnitude diagram \citep{col99b}.  In fourteen setups,
we obtained 52 spectra: the twenty-eight intended 
targets, a further twelve LMC red giants (one
of which was contaminated by strong TiO lines), one blue main-sequence 
star, and eleven red foreground dwarfs.  The foreground dwarfs were
identified and rejected based on their radial velocities.

\setcounter{figure}{0}
\begin{figure*}[t]
\centerline{\hbox{\epsfig{file=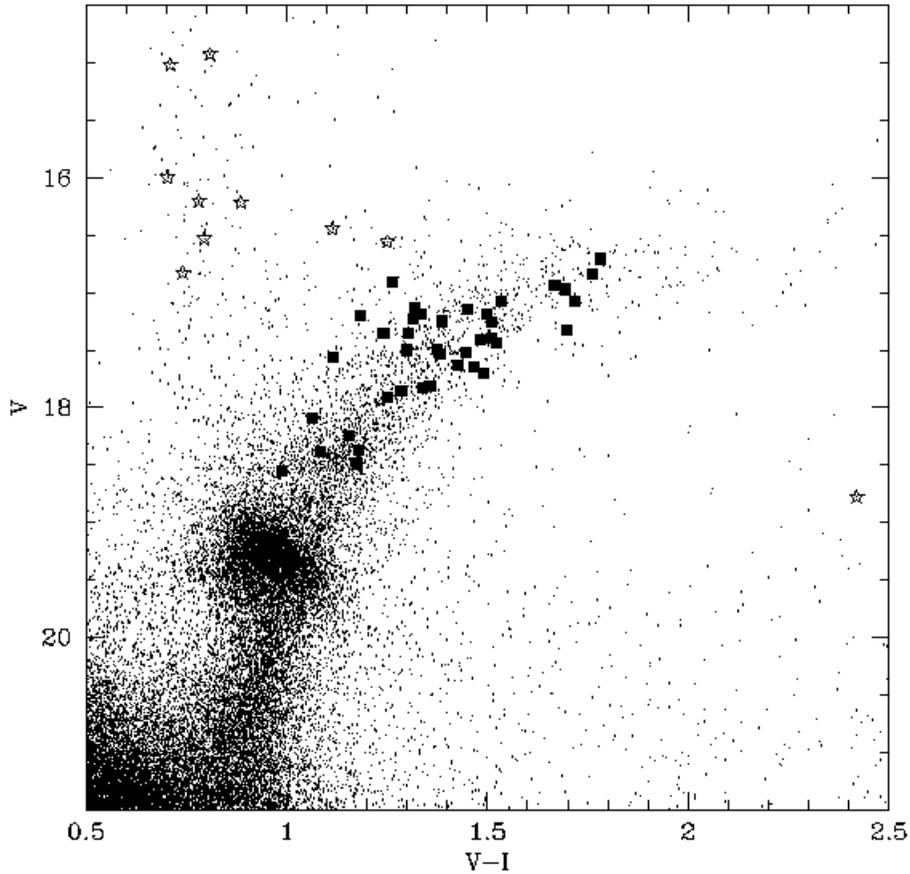,height=5.0in}}}
\caption{CMD of the Disk 1 field obtained from the CTIO 1.5m telescope.
39 LMC member giants for which abundance measurements were made are
marked with solid squares.  Excluded foreground stars are marked as 
open stars, as is star 36, afflicted with strong TiO lines. \label{speccmd}}
\end{figure*}

The bona fide LMC giants are listed in Table \ref{datatab}.  Thirty-eight
of the targets were easily matched to stars in the USNO-A astrometric
reference catalog \citep{mon98}, and for these stars the USNO number and
coordinates are given in columns 2--4.  Columns 5 and 6 give the summed
equivalent width of the two strongest lines of the \caii\ triplet 
and the associated 1-$\sigma$ error,
in \AA ngstroms; columns 7 and 8 give the V magnitide and error of each
star, from Cole (1999); columns 9 and 10 respectively list the 
metallicity of the star according to the calibration in \S 2.3, and
the random error associated with the metallicity value.  
Column 11 divides the sample of red giants
into four subgroups for further analysis.  Group {\it i} stars lie fainter
than 18th magnitude; group {\it ii} stars are brighter than V = 18 and
bluer than V$-$I = 1.6; group {\it iii} stars lie between V$-$I = 1.6
and 1.85.  Additionally, five of the bluest stars in selected magnitude
ranges are classed into group {\it iv}.  These subdivisions will be 
discussed further in \S 2.4 below.  Star 36, strongly affected
by TiO lines, is excluded from further analysis.

Our selected targets span the width of the RGB in the magnitude range
16.7 $\leq$ V $\leq$ 18.5.  The distribution of targets in the CMD is
shown in Figure \ref{speccmd}.  Our selection process should not strongly bias
our sample against the detection of metal-poor stars; comparison to 
globular cluster fiducial giant branches \citep{dac90} shows that
roughly one-third of our RGB sample lies to the blue of the [Fe/H] 
= $-$1 globular cluster fiducial sequence.

In order to calibrate our metallicity scale to the system of 
\cite[b]{rut97a}, we also oberved a sample of giants and red 
clump stars in five Galactic globular clusters and the 
open cluster M67. The clusters span a factor of $\sim$3 in age, and
two orders of magnitude in heavy element abundance.  The clusters,
their horizontal branch magnitudes, abundances, and calcium enhancements
are given in Table \ref{globref}. 
The individual calibrator stars, magnitudes, and equivalent widths are
given in Table \ref{globcal}; exposure times ranged from 30--300 seconds.

Bias exposures and dome flats were taken at the start of each night; sky
exposures to correct for fringing in the CCD were taken during evening and
twilight.  We observed the 13th magnitude B1 supergiant Sk $-$71$\arcdeg$ 11
on night 1 in order to estimate the effect of telluric absorption lines on a 
relatively featureless spectrum.  Because the data were taken during bright
time, through clouds, every spectrum included \caii\ lines from the reflected
and scattered sunlight.  However, we found that all atmospheric and solar
features subtracted out cleanly from the region around the redshifted 
\caii\ lines in the LMC target spectra.

The data were reduced with the IRAF {\tt ccdproc} and {\tt ctioslit} tasks, 
using the bias frames, overscan region, dome flats, and sky flats
to remove the instrumental signatures from the frames.  The sky fringes
showed a maximum peak-to-peak relative amplitude of $\approx$6\%, and 
were removed with little difficulty.  A number of bad columns were masked
off during the reductions, to prevent them from affecting the line
identification or measurement procedures.

Because most exposures contained at least two target objects, it was
necessary to identify the spectrum of each star interactively.  Each
spectrum was traced along the dispersion axis, using a linear fit where
possible.  Near the ends of the slit, it was necessary to use a cubic 
spline to trace and extract the spectra.  The stellar spectra were
extracted by interactively fitting a Gaussian shape to the stellar 
profile across the dispersion direction, 
placing sky windows appropriately to avoid the wings of
neighboring stars and bad pixels.  Spectra were extracted using
an optimal weighting procedure; the spectra were then
rectified using spline fits to selected areas of the continuum that were
judged to be unaffected by atmospheric absorption or emission lines.

\subsection{Equivalent Widths and Abundances \label{ewsec}}

Rutledge et al. (1997a) summarize the various approaches to measuring the \caii\
triplet in the literature.  We have followed the procedure of Suntzeff et al. (1993),
using Gaussian fits to the line profiles of the two strongest lines
of the triplet ($\lambda \lambda$ 8542, 8662 \AA).  The weakest line is
excluded because its inclusion usually decreases the final signal-to-noise
of the summed equivalent width measurement \citep{arm91,rut97a}.  We followed
Suntzeff et al. (1993) and Olszewski et al. (1991) in adopting the continuum 
bandpasses of 
Armandroff \& Zinn (1988) in our measurements.  We used the IRAF task 
{\tt rvidlines}
to define the wavelength scale of each spectrum, and measured the
equivalent widths relative to the near-IR stellar pseudo-continuum
using {\tt splot}.  Photometry for the calibration stars was taken
from the literature; references are given in Table \ref{globcal}.
Photometry for the LMC target stars was taken from our own WFPC2/HST
data \citep{sme99,col99b}.

The equivalent widths for the calibrating cluster stars are given in 
Table \ref{globcal}, spanning the range 2.2 $\lesssim$ $\Sigma$W(Ca)
$\lesssim$ 6.1.  Of the 10 stars in our sample in common with
Suntzeff et al. (1993), we find their equivalent widths to be systematically
smaller than ours by 1.5\%, although this trend is largely masked
by the scatter of $\pm$0.22 \AA\ between the two samples.
We have 7 calibration stars in common with Olszewski et al. (1991);
because they included the $\lambda$8498 line in their measurements,
their equivalent widths are 26\% larger than ours, with a scatter of
$\pm$0.09 \AA\ around the mean relation.  Our calibration sample only
contained two stars in common with Rutledge et al. (1997a), both in 
M 68 (NGC 4590).  Those authors used a weighted sum of all three
triplet lines to compute $\Sigma$W(Ca), where we took a straight
sum of the two strongest lines.  The Rutledge et al. (1997a) weighting
scheme generally assigns a very low weight to the weakest line, and
this is reflected by the fact that for the two stars in common, 
$\Sigma$W(Ca)$_{R97a}$ = 1.05 $\pm$0.05 $\Sigma$W(Ca)$_{new}$.
We will show in the next section that we may safely use the Rutledge et al.
(1997b) metallicity calibration with our data.

Equivalent widths for the
LMC targets are given in Table \ref{datatab}; the red clump 
of the Disk 1 field is measured to lie at V = 19.26, with a
full-width at half-maximum of 0.29 mag \citep{col99c}.
Because the magnitude
error on an individual red clump star is $\approx$0.03 mag, 
the width of the clump is due to stellar evolutionary effects
and the mixture of stellar populations in the field.  We consider
the possible implications of these variations on our MDF
in section \ref{corsec}.

\subsection{Calibrating the $\Sigma$W(Ca)-[Fe/H] Relation \label{r97sec}}

The behavior of $\Sigma$W(Ca) is biparametric with surface gravity
and calcium abundance; Rutledge et al. (1997b) describe the procedure for
removing the signature of gravity from $\Sigma$W(Ca) in globular
clusters.  This correction is possible because on the giant branch
of a star cluster, there is a one-to-one relation between luminosity
and surface gravity: stars along the length of the giant branch have
nearly the same mass, but the more highly evolved stars have larger
radii and hence lower surface gravity.  For low-mass, metal-poor stars,
the relation between surface gravity and V magnitude is very close to
linear between the level of the horizontal branch and the RGB tip.

An easily measurable surrogate for the difficult-to-measure surface
gravity, and the one that is most frequently adopted \citep[and references
therein]{rut97b}, is the difference in magnitude between a target star
and the horizontal branch level, \vhb.  Using the \vhb\ method, it is 
possible to reduce the $\Sigma$W(Ca) to a number that depends only
on abundance, W$'$ (known as the reduced equivalent width).
The Rutledge et al. (1997b) result, based on the catalog
of Rutledge et al. (1997a) for globular clusters in the approximate range 
$-$0.3 $\gtrsim$ [Fe/H] $\gtrsim$ $-$2.3, is:

\begin{equation}
\label{deltav}
\mathrm{W'} = \Sigma \mathrm{W(Ca)} + 0.64 (\pm0.02)\,
 (\mathrm{V} - \mathrm{V_{HB}}).
\end{equation}

We adopt the abundance scale of Carretta \& Gratton (1997), which was shown by
Rutledge et al. (1997b) to scale linearly with W$'$ according to the relation

\begin{equation}
\label{metcal}
\mathrm{[Fe/H]_{CG97}} = -2.66 + 0.42\, \mathrm{W'}.
\end{equation}

Both the slope and zeropoint of the giant branch relation
between log g and M$_V$ are theoretically
predicted to depend on age and metallicity (see, e.g., the models
of Girardi et al. 2000).  Also, it remains uncertain over what range
of stellar parameters the linear relations between
surface gravity, metallicity, and $\Sigma$W(Ca) hold true
\citep{dia89,jor92}.  However, the methods espoused by 
Rutledge et al. (1997b) have been used for targets much more youthful and/or 
metal-rich than the calibrating globulars with apparent success.

For example, de Freitas Pacheco et al. (1998) obtained integrated optical spectra and
V$-$K colors
for five interemediate-age LMC clusters 
(NGC 1783, NGC 1978, NGC 2121, NGC 2173) and 
three ancient clusters (NGC 1466, NGC 2210, H 11) that had been
observed by Olszewski et al. (1991).  Within the errors of both methods,
the \caii\ triplet and the spectral indices yield similar 
abundances for the ancient clusters, as expected.  For the 
intermediate-age clusters, there is a large scatter between
the methods.  Because the integrated colors used by de Freitas Pacheco et al. (1998)
are susceptible to age-metallicity degeneracy, we shift their
measured abundances to the values that would have been obtained
if the cluster ages from color-magnitude diagrams had been
adopted.  We used the relation provided by de Freitas Pacheco et al. (1998), namely
$\frac{\partial [Fe/H]}{\partial ln \tau}\vert _{(V-K)}$ = $-$0.329.
Comparison of the abundances derived from the de Freitas Pacheco et al. (1998) paper
and those of Olszewski et al. (1991) for four intermediate-age clusters
shows that the \caii\ triplet values are lower than those derived
from spectral indices, but only by 0.1 $\pm$0.2 dex.  Despite 
expectations for the breakdown of the \caii\ method at young ages
and high abundance, we see reasonable agreement with other methods
for ages as young as $\approx$1 Gyr and abundances [Fe/H] $\lesssim$ $-$0.25
dex.

\begin{figure*}[t]
\centerline{\hbox{\epsfig{file=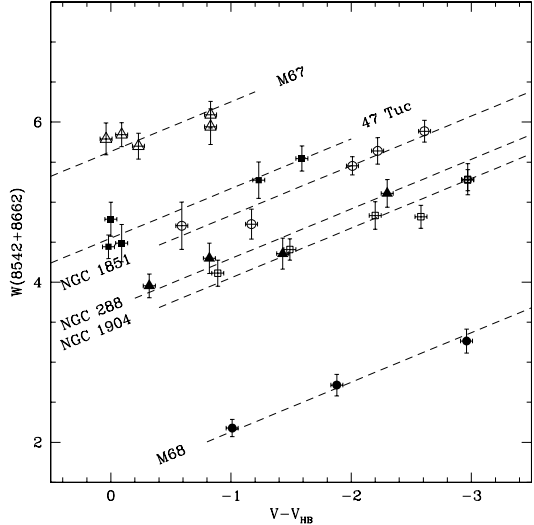,height=3.0in}}}    
\caption{Summed \caii\ equivalent widths plotted
against \vhb\ for stars in our calibrating clusters.  For each
cluster, the best-fit line with slope 0.64 is shown (see text).
\label{calvhb}}
\end{figure*}

The summed \caii\ equivalent widths are plotted as a function
of \vhb\ for the stars in our calibrating clusters in Figure 
\ref{calvhb}.  The best-fit lines of slope 0.64 (from Equation
\ref{deltav}) are plotted for each cluster.  Extrapolating the
lines back to \vhb\ = 0, we can derive values for W$'$ for the
clusters.  Although we have used a slightly different technique
than Rutledge et al. (1997a) for measuring the summed equivalent
width of the calcium triplet, we find essentially identical values
of W$'$ for the five globular clusters common to the two samples:

\begin{equation}
\label{wprime}
\mathrm{W'}_{new} = 1.02\, \mathrm{W'}_{R97a} + 0.002;
\end{equation}

the difference between the Rutledge et al. (1997a) system and
our own is much smaller than the mean error of the mean in W$'$,
which is $\pm$0.11 \AA.  We can test this by using our values of
W$'$ to rederive an abundance calibration:

\begin{equation}
\label{metcal2}
\mathrm{[Fe/H]_{CG97}} = -2.656 + 0.402\, \mathrm{W'}.
\end{equation}

\noindent This is in excellent agreement with Equation \ref{metcal},
showing that the differences in calculation of $\Sigma$W(Ca) will
not affect our abundance analysis. Such behavior was also found by 
Smecker-Hane et al. (1999b), who note that the equivalence will not
generally hold true in cases where the instrumental resolution and
scattered light properties of the spectra differ strongly from 
that of the Rutledge et al. (1997a) setup.

Equation \ref{metcal2} was obtained {\it excluding M67 from the fit}.
According to this fit, M67 should have an abundance of $-$0.3 dex,
incompatible with the known [Fe/H] = $-$0.06 $\pm$ 0.07 
\citep{nis87}.  The discrepancy is understandable because no
cluster as metal-rich as
[Fe/H] = 0 exists in the calibrating sample of Rutledge et al. (1997a),
and theory leads us to expect a breakdown of the linear relation
between W$'$ and [Fe/H] for clusters of solar abundance 
\citep{jor92}.  

\begin{figure*}[t]
\centerline{\hbox{\epsfig{file=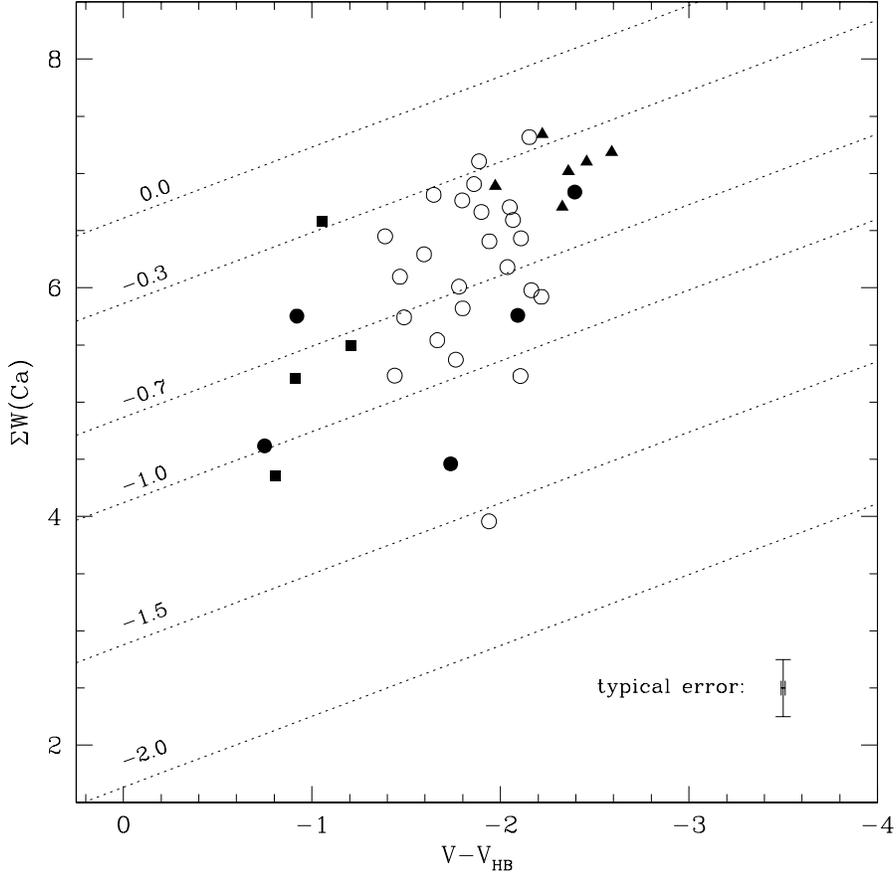,height=5.0in}}}
\caption{Summed \caii\ equivalent widths plotted
against \vhb\ for LMC target stars.  A typical errorbar is shown at 
lower right.  Dotted lines show constant [Fe/H], labelled at left.
Symbol key to subgroups: $\blacksquare$ = faint; $\circ$ =
main; $\blacktriangle$ = tip; $\bullet$ = blue.
\label{lmctargs}}
\end{figure*}

The summed \caii\ equivalent widths for the LMC target stars
are plotted against \vhb\ in Figure \ref{lmctargs}.  Errorbars
are omitted for clarity, but a typical 1-$\sigma$ error is
plotted in the lower right corner.  Dotted lines denote isofers
derived from equations \ref{deltav} and \ref{metcal2}.  The four
subgroups of Table \ref{datatab} are plotted with different symbols.
Group {\it i}, the faint stars, are solid squares; stars on the main
body of the RGB (group {\it ii}) are plotted as open circles; the
stars near the tip of the RGB (group {\it iii}) are closed triangles;
and group {\it iv}, the bluest stars, are closed circles.

\subsection{The Spectroscopic Metallicity Distribution Function \label{smsec}}

\begin{figure*}[t]
\centerline{\hbox{\epsfig{file=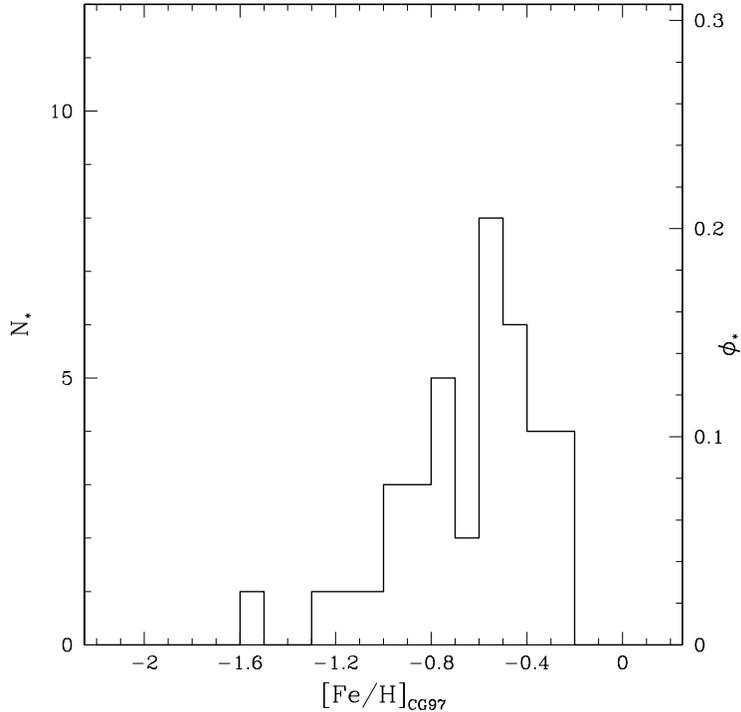,height=4.0in}}}
\caption{MDF, derived from the \caii\ triplet,
for 39 LMC red giants.  [Fe/H] is calibrated following Rutledge et al. (1997a)
to the system of Carretta \& Gratton (1997) (see text for details).  Labels show
true number (N$_{\star}$) and fractional distribution ($\phi _{\star}$).
The mean abundance is [Fe/H] = $-$0.642.
\label{ca2hist}}
\end{figure*}

The abundance histogram for our spectroscopic sample is shown in 
Figure \ref{ca2hist}. 
The random, 1-$\sigma$ errors for individual stars range
from 0.10--0.17 dex, with a mean error of $\pm$0.13 dex.  [Fe/H] values
and the associated random error for each star are given in Table
\ref{datatab}.
The sharpness of the peak in the [Fe/H] = $-$0.55 $\pm$0.05 bin of 
Fig. \ref{ca2hist} is partially an artifact of the positioning of the 
bin edges.  
The dominant error terms are due to {\it 1)} errors
in the equivalent width, {\it 2)} scatter in the calibration relation,
and {\it 3)} uncertainty in the location of the red clump.  These are
random errors and do not take into account systematics due to age
or [Ca/Fe] variations.

The median [Fe/H] = $-$0.571 $\pm$0.037, while the mean of the sample is
[Fe/H] = $-$0.642 $\pm$0.022.  The maximum metallicity is [Fe/H] = 
$-$0.251 $\pm$0.112, and the most metal-poor star in the spectroscopic
sample has [Fe/H] = $-$1.548 $\pm$0.127.  The distribution can be 
characterized by a 1-$\sigma$ dispersion about the mean of $\pm$0.30 dex,
or by the locations of the 1st and 3rd quartile points at $-$0.828 dex
and $-$0.413 dex, respectively.

\begin{figure*}[t]
\centerline{\hbox{\epsfig{file=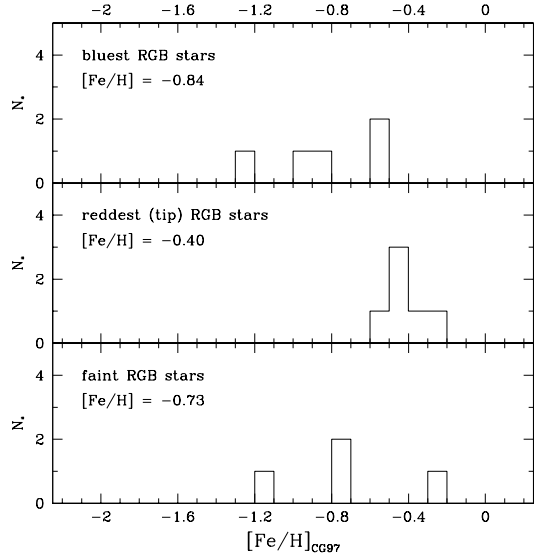,height=3.0in}}}
\caption{Abundance histograms for the three
minority subgroups in our sample, with their respective mean abundances.
The results bear out theoretical predictions for a correlation of giant
branch color with metallicity.  The faint subgroup may be more metal-poor
than the sample as a whole-- see text for discussion.
\label{grphist}}
\end{figure*}

The abundance distributions in the minority subgroups from Table 
\ref{datatab} are plotted in Figure \ref{grphist}.  The top panel
shows the abundances of the five bluest RGB stars; the mean is 
0.2 dex lower than the mean of the sample as a whole.  In contrast,
the stars near the RGB tip (redder than V$-$I = 1.6) are shown in 
the middle panel of Fig. \ref{grphist}.  Their mean abundance is
0.24 dex higher than the mean of the sample.  The lower panel
of Figure \ref{grphist} shows that the giant branch between the 
red clump and V = 18 contains stars spanning the full range of abundances
in this field.  The lower mean abundance is not statistically significant,
given the small sample size; if such a trend persists to larger 
spectroscopic samples, then we would be forced to examine the effect
of magnitude bias on the derived MDF.

The color-abundance 
correlation readily lends itself to interpretation in the framework
of stellar evolution theory: as expected, the bluer stars are, in 
general, more metal-poor than the redder stars.  Note that this
trend can be largely erased for field star samples if the epoch of
major star-formation persists for a timescale longer than the typical
chemical enrichment timescale (e.g., Aparicio et al. 1996, Da Costa 1998).  This is 
because young, metal-rich red giants can have very similar
colors to ancient, metal-poor red giants.  The implication from the
top two panels of Figure \ref{grphist} is that a mean age-metallicity
relation can well-characterize the evolution of the LMC disk, in this
field (c.f., Holtzman et al. 1999).

\subsubsection{Corrections to the W$'$-[Fe/H] Relation \label{corsec}}

For better or for worse, the Large Magellanic Cloud is not a globular
cluster.  While this has proved a boon for studies of stellar evolution
and galactic structure, it introduces several complications to the attempted
derivation of its metallicity distribution function.   The issues of 
age and $\alpha$-enhancement were discussed in the context of the \caii\
triplet by Da Costa \& Hatzidimitriou (1998). 
We consider these, as well as the issue of AGB
star contamination, here.

There are known to be a large number of AGB stars present in the LMC field,
relative to those in a truly ancient population (e.g., Frogel, Mould \& Blanco 1990).
We find that synthetic CMDs show that the number of early-AGB stars lying hidden
in the broad RGB may be as high as 30\% of the total number of RGB$+$AGB stars.
The calibration of \caii\ equivalent width vs. [Fe/H] is tuned to 
first-ascent red giants, and so may be systematically biased when
applied to an intermediate-age mixture of RGB and AGB stars.  We used 
the NLTE line-formation results of J\o rgensen et al. (1992), combined with the
theoretical stellar evolution models of Girardi et al. (2000) to model this
bias.  Among stars of a constant age and metallicity, we calculate
that a red giant at given V magnitude will have a \caii\ equivalent
width $\approx$0.04 \AA\ larger than an AGB star at the same magnitude.
This translates directly to an error in [Fe/H] of $-$0.02 dex, much smaller
than the typical error in abundance of $\pm$0.1--0.2 dex.  Even if a third
of our measured stars are AGB stars, the derived mean abundance will be 
virtually unaffected\footnote{The Str\"{o}mgren photometric indices, being insensitive
to surface gravity, will be even less influenced by the presence of AGB stars
or even red supergiants.}.

The second possibly serious source of error in our analysis is the
use of \vhb\ to account for the effect of surface gravity on the
\caii\ equivalent widths.  Because theory predicts an exponentially
increasing dependence of $\frac{\partial \Sigma W(Ca)}{\partial logg}$
on [Fe/H] \citep{jor92}, this problem will be most severe for the
high-metallicity end of our sample, precisely where the cluster 
sample of Rutledge et al. (1997a) becomes severely incomplete.  

We can break down the dependencies of $\Sigma$W(Ca) on various 
stellar parameters, and it is found that the slope of Equation
\ref{metcal} decreases with increasing metallicity.  Additionally,
the slope of Equation \ref{deltav} depends not only on the physics
of \caii\ line formation, but on the slope of the giant branch.
The zeropoint of Equation \ref{metcal} depends on the surface gravity
of an RGB star at the level of the red clump/horizontal branch.
Combining the theoretical models of Girardi et al. (2000) and
J\o rgensen et al. (1992),
we find that linear equations like Equations \ref{deltav} and \ref{metcal}
are appropriate for abundance determinations of {\it old, metal-poor stars};
Rutledge et al. (1997a) have observationally shown this to be true for globular 
clusters.  However, for stars approaching solar metallicity, and younger
than a few billion years, theory predicts a gradual lessesning of 
sensitivity of [Fe/H] to $\Sigma$W(Ca), until the \caii\ triplet eventually
becomes nearly useless as a metallicity indicator above a certain 
abundance.  Because of the gravity-dependence, the exact value of 
the threshold [Fe/H] depends on the stellar mass, and hence age.

\begin{figure*}[t]
\centerline{\hbox{\epsfig{file=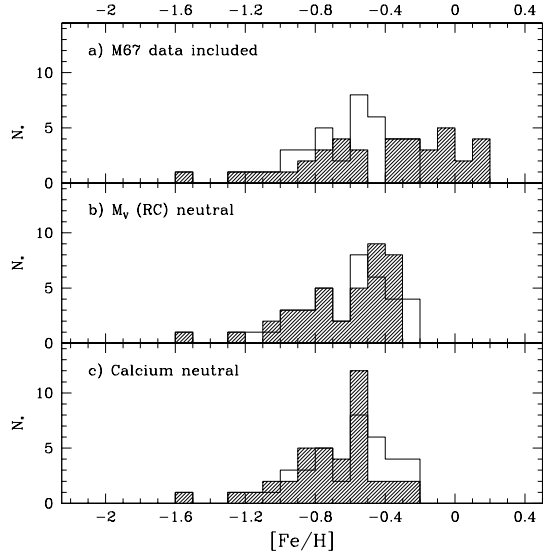,height=3.0in}}}
\caption{Estimated corrections to the spectroscopic 
MDF for three different systematic effects.  The corrected MDFs are 
plotted as shaded histograms, on top of the MDF from Fig. \ref{ca2hist}.
The panels show the effects of: {\it a)} adding M67 back to the 
calibrating sample; {\it b)} considering the variation of red clump
magnitude with age and metallicity; {\it c)} considering the difference 
in [Ca/Fe] between the LMC and the calibrating clusters.
\label{comphist}}
\end{figure*}

We expect the M67 data to deviate from the Rutledge et al. (1997b) calibration
line, and indeed this behavior is seen in Figure \ref{calvhb}.  If
we put aside our theoretical expectations for the moment, and switch to
a bilinear calibration of [Fe/H] vs. W$'$, including M67 in the new fit,
we obtain the MDF shown in Figure \ref{comphist}a.  The large equivalent
widths of the bright LMC giants have caused their abundances to be
shifted upward by $\lesssim$0.5
dex, to nearly twice solar metallicity.  Such high values are unprecedented
(e.g., Jasniewicz \& Th\'{e}venin 1994),
and clearly show the nonlinear dependence of W$'$ on [Fe/H] in this 
age-metallicity regime.  In fact, we have already seen that the \caii\
triplet applied to LMC clusters as young as $\approx$ 1 Gyr yields values
within $\approx$ 0.1 dex of values determined using spectrophotometric
indices (q.v. \S \ref{r97sec}).  Fortunately, the LMC is not as metal-rich
as M67 and so the \caii\ triplet retains sensitivity for correspondingly
large values of stellar surface gravity (i.e., younger ages).  The effect
shown in Figure \ref{comphist}a shows an extreme case that certainly
does not apply to our data; however the drawing upwards of the 
high-metallicity half of our sample by $\lesssim$0.1 dex is a real
possibility.  Therefore, we have undertaken to combine new measurements
of the \caii\ triplet in open and globular clusters with theoretical
stellar atmosphere models to remove this uncertainty from future
studies of LMC field stars.

Another way in which age and metallicity influence our analysis is in
the identification of V$_{HB}$ (or V$_{RC}$) itself;
in a star cluster this is relatively
straightforward, but in a composite stellar population, the identification
of a specific red giant star with a specific HB (or red clump)
level is problematic.
The 1-$\sigma$ magnitude width of the red clump in our field is 
$\pm$0.12 mag, which adds a random error of $\pm$0.03 dex to our
abundance determinations.  However, the red clump magnitude is 
expected to vary in a regular way with age and metallicity, and
so the use of a mean V$_{RC}$ for all stars can produce systematic 
errors.

\begin{figure*}[t]
\centerline{\hbox{\epsfig{file=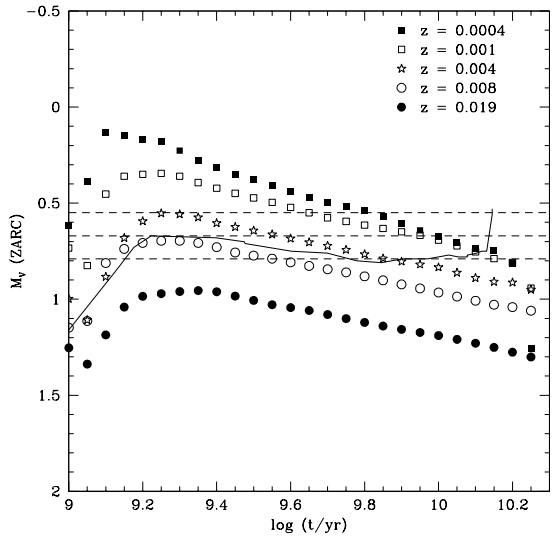,height=3.0in}}}
\caption{Variation of zero-age red clump magnitude
with age and metallicity, from the theoretical models of Girardi et al. (2000).
The thick solid line denotes the predicted M$_V$, given a sample
age-metallicity relationship for the LMC.  Dashed lines show the 
observed M$_V$ (RC) if the distance modulus is assumed to be 18.5,
with the 1-$\sigma$ width of $\pm$0.12 mag included.
\label{rcfade}}
\end{figure*}

We show the predicted variation of red clump magnitude as a function
of age and abundance in Figure \ref{rcfade}; models are from Girardi et al. (2000).
The magnitude of the zero-age red clump (ZARC) is shown for simplicity.
We have taken the age-metallicity and age-[O/H] relations of
Pagel \& Tautvai\v{s}ien\.{e} (1998),
and derived age-[M/H] relations \citep{sal93} for comparison to the 
stellar evolution models of a given abundance.  The thick solid line
shows the predicted variation of M$_V$ (ZARC) in the LMC disk.  For
comparison, we show the value of M$_V$ (RC), given our observed value
of V$_{RC}$ = 19.26 $\pm$0.12 and assuming a distance modulus of 18.5 mag, as 
dashed lines.  The agreement with theory is encouraging.  Figure 
\ref{rcfade} shows that, by using an average value of V$_{HB}$ in
our analysis, we may have overestimated the metallicity of both
the high- and low-metallicity ends of the MDF, while the stars near
the peak of the MDF may have had their abundances underestimated by
a small amount.  

If a correction for clump magnitude is made, the highest abundance
stars are decreased in [Fe/H] by as much as 0.1 dex.  The 
correction does not exceed $\pm$0.03 dex for any star more metal-poor than 
[Fe/H] $\approx$ $-$0.4 in this model.  The ``clump neutral'' MDF is
shown in Figure \ref{comphist}b.

A third potential source of systematic error is the known enhancement
of $\alpha$-elements in the calibrating sample of Galactic globular
clusters.  Column 5 of Table \ref{globref} shows the measured [Ca/Fe]
ratio for the globulars we have observed.  There is a clear anti-correlation
of [Ca/Fe] with [Fe/H] (e.g., Gilmore \& Wyse 1991; Carney 1996; McWilliam 1997).
Tinsley (1979)
first explained this trend as a result of the very different timescales
for Type II supernovae ($\alpha$ + Fe producers) and Type Ia supernovae
(Fe producers).  Thus the [$\alpha$/Fe] ratio should be controlled by
the relative strength and duration of star-formation epochs vs. quiescent
periods in galactic evolution, with additional influence from such factors
as could vary the Type Ia/Type II SNe ratio.  

If the LMC shared the same
\sfh\ and chemical evolution history as the Milky Way, measuring
[Fe/H] from lines of calcium would not pose a problem.  However,
this is not the case. 
The present-day LMC is thought to be mildly $\alpha$-deficient
relative to the Solar neighborhood (e.g., Hill et al. 1995 find 
[Ca/Fe] = $+$0.11 for [Fe/H] = $-$0.24), but its [$\alpha$/Fe] 
ratio is expected to change with age in a non-monotonic way \citep{pag98}.
Because the LMC's
star-formation may have been dominated by an intermediate-age burst (e.g.,
Stryker 1984), there is likely to be a peak in [$\alpha$/Fe] for
stars aged slightly less than the burst age, with lower values for
both younger and older ages.

\begin{figure*}[t]
\centerline{\hbox{\epsfig{figure=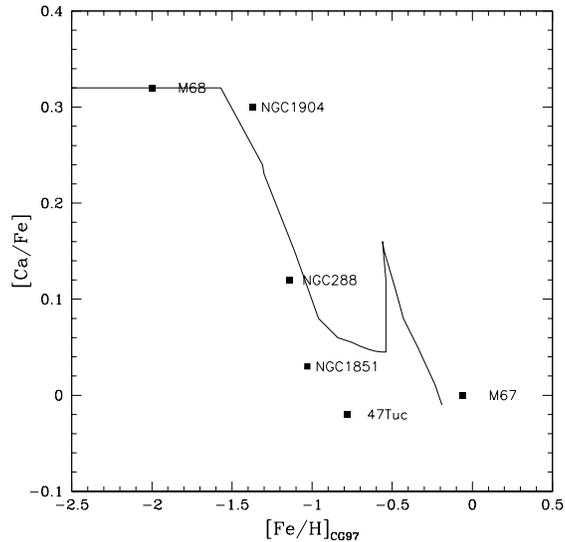,height=3.0in}}}
\caption{Variation of [Ca/Fe] with [Fe/H] for the
calibrating clusters (solid squares), and a model for the LMC
(solid line).
\label{cafefig}}
\end{figure*}

We can make a zero-order correction for the effects of
varying [Ca/Fe], although this will rest on assumptions about
the \sfh\ and the particulars of the adopted enrichment model.
Pagel \& Tautvai\v{s}ien\.{e} give a theoretical relation between [Ca/Fe] and 
[Fe/H], for an \sfh\ model similar to that proposed by Holtzman et al. (1997).
We plot the theoretical [Ca/Fe] vs. [Fe/H] curve \citep{pag98}
in Figure \ref{cafefig}, along with the [Ca/Fe] values from the
calibrating clusters.  The globular cluster values are included in 
Table \ref{globref}; they are taken from Carney (1996) and span 
the range $+$0.32 $\geq$ [Ca/Fe] $\geq$ $-$0.02
We estimate ``calcium neutral'' [Fe/H] values by correcting the
values inferred from the \cite{rut97b} calibration, according to
the difference in [Ca/Fe] between the globular clusters and the
LMC model.

The adopted corrections are hidden within a $\pm$0.04
dex scatter, below [Fe/H] $\approx$ $-$1.1; in the range $-$1.1
$\lesssim$ [Fe/H] $\lesssim$ $-$0.55, the correction gradually 
increases to a value of $-$0.07 dex.  Where the few-Gyr ``burst''
in the LMC's star-formation rate occurs, the correction jumps
to $-$0.17 dex, due to the sudden increase in the Type II SNe rate, and 
then falls rapidly to negligible values for the most metal-rich
stars.  The ``calcium neutral'' MDF is shown in Figure 
\ref{comphist}c.  This figure shows that the discontinuity in 
[Ca/Fe] at the metallicity of the intermediate-age burst may have
caused us to overestimate the metallicity of the upper quartile 
of our sample.  The low-metallicity tail is barely changed, although
pulled to slightly lower values by $\lesssim$0.07 dex.  The 
quantitative values of the adopted correction are highly susceptible
to systematic error stemming from the exact timing and duration of 
the burst, but the qualititative behavior, that of strengthening the
peak of the MDF at the expense of the high-metallicity component,
should be robust.  However, the effect is almost
certainly hidden within the large scatter of [Ca/Fe] among 
moderate-metallicity stars in the LMC 
\citep[and references therein]{pag98}.  This scatter is to be
expected, because it requires multiple supernovae to enrich a
star-forming region to [Fe/H] $\approx$ $-$1.  Variations in 
[$\alpha$/Fe] will likely increase with increasing [Fe/H], an 
effect which is seen in the population of Galactic globular clusters
\citep{kra94}.

The posible systematic effects on our analysis, and their magnitude,
are summarized here:

\begin{enumerate}

\item The contamination by AGB stars induces a very small 
($\approx$ 0.02 dex) bias to smaller abundance for $\approx$30\%
of the stars.

\item The stars above [Fe/H] $\approx$ $-$0.4 may be biased
too low by up to $\sim$ 0.1 dex due to the nonlinear dependence
of [Fe/H] on log g and $\Sigma$W(Ca).

\item The peak of the MDF is probably sharper than observed
in Figure \ref{ca2hist}, and the stars above [Fe/H] $\approx$
$-$0.4 may be biased too high by up to $\approx$ 0.1 dex due
to the variation of clump magnitude with age and abundance
(note that this effect works against item \#2).

\item The peak of the MDF may be much sharper than observed
in Figure \ref{ca2hist}, and the stars above [Fe/H] $\approx$ 
$-$1 may be biased too high by $\lesssim$0.1 dex, due to the
changing ratio of [Ca/Fe].  This systematic is the least well
constrained of any, and is probably completely hidden by the
scatter in [Ca/Fe] at a given [Fe/H].

\item None of the systematic effects we have noted in this 
section are capable of populating the low-metallicity tail
of the MDF-- note the similarity of Figure \ref{comphist}a,b,c
for [Fe/H] $\lesssim$ $-$1.  This fact will become important
in our discussion of spatial variations of the MDF (\S \ref{varsec}).

\end{enumerate}

\section{Str\"{o}mgren Photometry \label{phsec}}

Intermediate-band photometry represents a useful compromise between
the photon-gathering ability of wide bandpass filters, and the increased
information contained in a high quality spectrum.  The Str\"{o}mgren 
system was designed to match bandpass windows to strong and distinctive
features of a normal stellar spectrum.  In this way, astrophysically 
meaningful color indices of high utility can be formed \citep{str63}.

For late-type (G--M) stars, the primary Str\"{o}mgren filters are $v$
($\bar{\lambda}$ = 4118 \AA, $\delta \lambda$ = 146 \AA), $b$ 
($\bar{\lambda}$ = 4697 \AA, $\delta \lambda$ = 196 \AA), and $y$
($\bar{\lambda}$ = 5497 \AA, $\delta \lambda$ = 241 \AA).  The $y$
filter transmits no strong spectral features, and is equivalent to
a narrower version of Johnson V.  The $b$ filter is placed just
longward of the wavelengths where metal-line blanketing becomes
important, thus $b-y$ is a good temperature indicator.  The $v$
filter lies just longward of the Balmer jump; a stellar spectrum 
in this wavelength region is rife with metal lines.  The strongest
photospheric lines in $v$ are due to iron peak elements and cyanogen
bands \citep{bel78}.
The index $m_1$ $\equiv$ ($v-b$)$-$($b-y$) 
therefore measures the stellar metal abundance by gauging the
depression of $v$-band flux relative to the baseline established
by $b-y$ \citep{str63, cra70}.  The existence of CN-anomalous stars
presents a possible complication to the use of the Str\"{o}mgren filters
\citep{ant95}; the effect is expected to be minor for the 
LMC field stars (see \S 3.4.1), but may contribute to a
metal-dependent offset between abundances derived from the
\caii\ triplet and Str\"{o}mgren photometry (\S 4.1).

A useful feature of the Str\"{o}mgren system is that 
$(\frac{\partial (b-y)}{\partial log g})\mid_{T,z}\;
\approx (\frac{\partial m_1}{\partial log g})\mid_{T,z}\;
\approx 0$ \\ \citep{bel78, gre92, kur99}.  Because of this fact,
abundance estimates using Str\"{o}mgren colors will be much 
less susceptible to the dreaded age-metallicity degeneracy.
The disadvantage is that we will be forced to subtract the 
contribution of foreground, Galactic dwarf stars from our
color-metallicity plots (\S \ref{fgsec}).  This stands in 
contrast to a spectroscopic study, where radial velocities
serve to reject Galactic interlopers.

Isofers (lines of constant metallicity) in the ($b-y$, $m_1$)
diagram are well-approximated by straight lines within a
certain color range \citep{bon80}. Richtler (1989) derived
metallicity calibrations for G--K supergiants in the
($b-y$, $m_1$) diagram.  The calibration was extended by
Grebel \& Richtler (1992) to red giants, for the color range 
0.4 $\leq$ $b-y$ $\leq$ 1.1, corresponding to a
temperature range 3600 K $\lesssim$ \teff\ $\lesssim$ 6000.
This is well-matched to the temperature range of red clump
and red giant stars in the LMC.  Because the Grebel \& Richtler
(1992) calibration
was largely based on field stars in the Solar neighborhood,
it was only weakly constrained at low abundance; this situation
has recently been remedied with observations of globular 
clusters in the Str\"{o}mgren system by Hilker (2000).  The 
resulting data have greatly improved the accuracy of the
color-metallicity calibration for [Fe/H] $\lesssim$ $-$1.

\subsection{Observations, Reductions \& Photometry \label{orpsec}}

We obtained Str\"{o}mgren $vby$ images of four fields in the LMC 
disk and bar during the nights 1997 Nov. 28 -- Dec. 1, using the 
CTIO 1.5-meter
telescope and Cassegrain focus CCD imager.  The coordinates
of each field, names of nearby star clusters, and references to
deep WFPC2 CMD studies are given in Table \ref{fields}.  To emphasize
the wide range of environments our study will probe, all four fields
are listed, together with their radial distances from the center of 
the LMC.  We will consider only the Disk 1 data for the remainder
of this paper, in order to make the comparison of abundances
obtained via the Str\"{o}mgren method to those obtained using 
the \caii\ triplet.

The CCD was a SITe 
2048$^2$ chip with 24 $\micron$ pixels; in combination with the 
f/7.5 secondary mirror, this yielded a pixel scale of 0$\farcs$435
pixel$^{-1}$ and a field of view 14$\farcm$8 on a side.  Four 
amplifiers were used for faster readouts, which resulted in 
mild, quadrant-dependent variations in readnoise and gain;
the averages were 
$\sigma _{RN}$
= 1.4 $\pm$0.1 ADU and G$_{inv}$ = 3.0 $\pm$0.2 e$^{-1}$/ADU.  The
chip specs gave a quoted linearity of 4\% at 56000 ADU; tests prior
to the observations confirmed this, showing linearity better than 1\%
to a level of at least 49000 ADU.  

Observations were taken during
dark time, with new moon occurring near the midpoint of the run.  
Conditions were photometric throughout, with the exception of the middle
third of the last night of the run.  The seeing, measured by the
FWHM of a 2-dimensional Gaussian fit to the point-spread function of
isolated stars, was typically in the range 1$\farcs$3--1$\farcs$4 for
$v$ and $y$ images. The image quality in $b$ was frequently $\approx$20\% worse 
than in $vy$.  The image quality in long exposures was slightly
worse than in the short exposures of photometric standards,
and represents a convolution of the true, atmospheric seeing, and
tracking errors over the 10--20 minute object exposures.
The observing log is given in Table \ref{obslog1}.

Photometric standards from the list of Richtler (1990) were observed
throughout each night, spanning the range of airmasses encountered
in our LMC observations.  
The total exposure
times in in this field were t($v$,$b$,$y$) = (4$\times$1300, 2$\times$1300,
2$\times$1200) seconds.  The exposure times were chosen to permit high
signal-to-noise photometry for red giants down to the level of the
red clump at ($v$,$b$,$y$) $\approx$ (20.7,19.9,19.3).  

The images were reduced within IRAF\footnote{IRAF is distributed by 
the National Optical Astronomy Observatories, which are operated by
the Association of Universities for Reseach in Astronomy, Inc., under
cooperative agreement with the National Science Foundation.}, using 
standard tasks within the CTIO {\tt ared} package to correct for the zero
level and bias variations of the four-amplifier CCD.  The data were
flatfielded using dome flats to correct for the pixel-to-pixel 
response variation and smoothed twilight sky flats to correct for the
illumination pattern.  Additional processing was performed
on exposures shorter than 20 seconds to correct for shutter timing
differences across the field \citep{ste89}.

Photometry was performed using the DAOPHOT/ALLSTAR \citep{ste87} suite
of Fortran programs, together with ALLFRAME \citep{ste94} and associated
calibration routines (Stetson 1997, private communication).  These 
programs are designed to make empirical, parametric fits to the 
point spread function (PSF) of isolated stellar images,
find the best-fitting position and amplitude of the PSF of each 
star in a crowded field, and attempt to remove the stellar flux from
the image, in an iterative process.  The PSF in our images varied
cubically with chip position.  We refined our PSF starlist and 
rederived increasingly accurate PSFs with each iteration, until
the residuals in the star-subtracted images were indistinguishable
from the intrinsic noise in the images.

\begin{figure*}[t]
\centerline{\hbox{\epsfig{figure=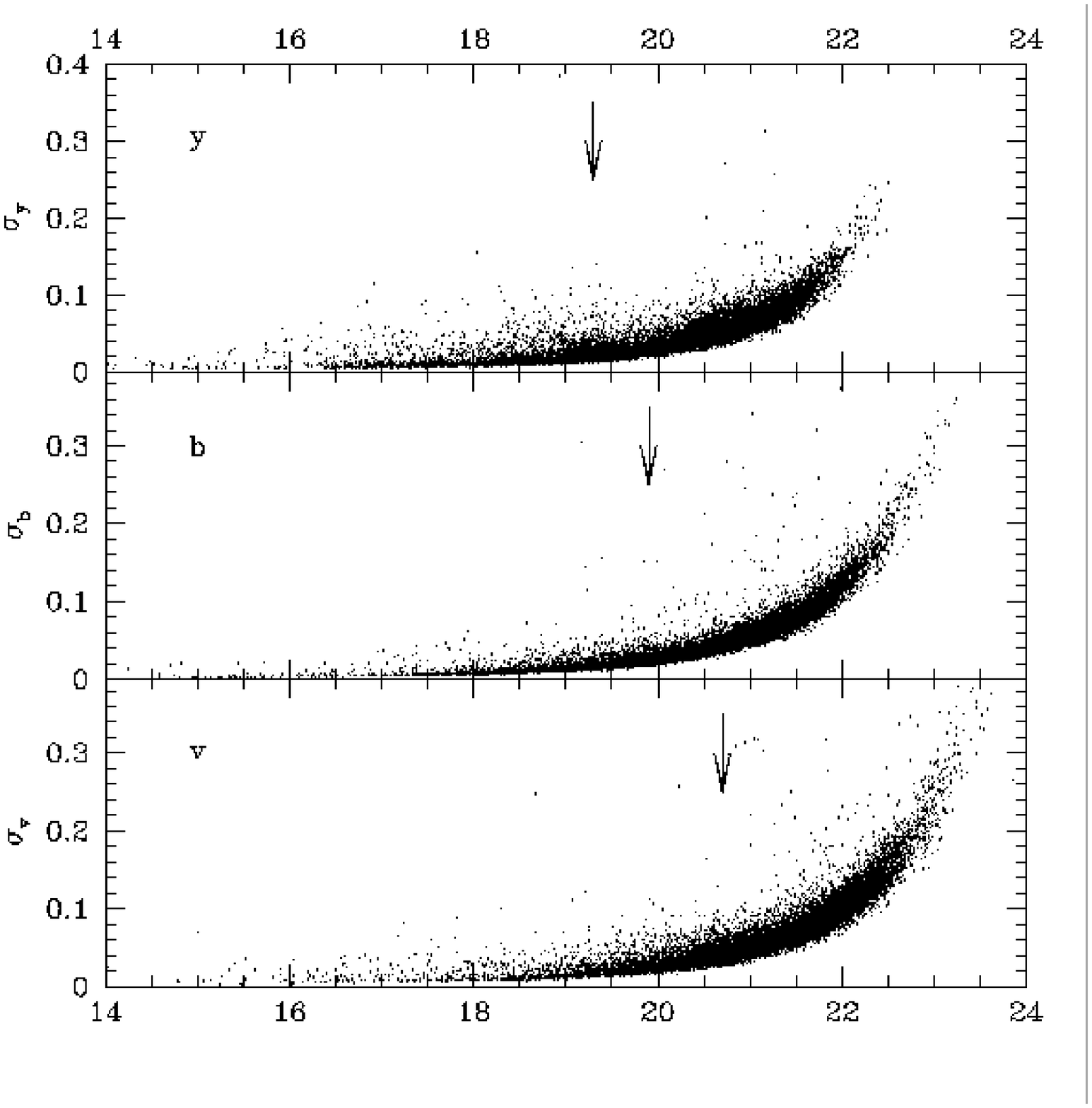,height=3.0in}}}
\caption{Photometric errors returned by ALLFRAME as a
function of stellar magnitude in each Str\"{o}mgren band.  Vertical 
arrows mark the position of the red clump in each band. 
\label{errfig}}
\end{figure*}

The ALLFRAME program improves upon the precision of measurements
derived from a single exposure by simultaneously analyzing
the $N$ exposures in different filters and solving for a
single ($x,y$) position and $N$ magnitudes per star and the
coefficients that describe the geometric transformation between images.
The star-subtracted images were then combined into a median image,
permitting the identification of faint stars that had previously fallen
below the detection threshold.  These faint stars were added to the 
master starlist and ALLFRAME was repeated.  This procedure greatly
extends the depth and precision of photometry.  The final distribution
of photometric errors with magnitude for each band is given in 
Figure \ref{errfig}; for reference, the magnitudes of the red HB clump
are marked with arrows.

Stellar magnitudes are computed
by the application of aperture corrections based on the measured
fluxes of bright, isolated stars in each image; the final magnitudes are intended
to represent the best possible approximation to the flux a given
star would contribute if it were observed with an infinitely large
aperture in an otherwise empty image with zero background.  For bright
stars, the uncertainty in the aperture correction, $\approx$2--3\%, 
dominates the photon noise.

For calibration we observed stars in the 
Sk $-$66$\arcdeg$80 (5 stars),
NGC 2257 (4 stars), and NGC 330 (6 stars) fields
\citep{ric90}.  These stars span
the color ranges 0.026 $\leq$ $b-y$ $\leq$ 1.138 and 0.040 $\leq$
$m_1$ $\leq$ 0.469.  Each star in NGC 330 was measured twice in $vby$ 
as well as in Johnson V;  NGC 2257 was observed three times; because
the Sk $-$66$\arcdeg$80 field was located just $\approx$5$\arcdeg$ 
north of our Disk 1 field, we were able to observe it five times during
the night of 1997 Nov 28.  We transformed observed colors to the standard
system using linear equations for zeropoint, airmass and color terms.
The rms scatter of standards about the fit, given here as 
$\sigma$, was $\approx$1--2\%:

\begin{eqnarray}      
y = y_{inst} + 3.832 + 0.102\, (b-y) - 0.171\, X\, \, \, \, 
\, \, \, (\sigma = \pm0.013), \\
b = b_{inst} + 4.004 + 0.173\, (b-y) - 0.232\, X\, \, \, \, 
\, \, \, (\sigma = \pm0.017), \\
v = v_{inst} + 4.438 + 0.421\, (b-y) - 0.326\, X\, \, \, \, 
\, \, \, (\sigma = \pm0.020),
\end{eqnarray}

The scatter around the calibration line was higher
for stars near the western horizon than near the eastern, for a given airmass.
An examination of the residuals to our photometric solution indicated
a problem with the star Sk $-$66$\arcdeg$82: its measured V magnitude was
consistently 1 $\pm$ 0.02 mag brighter than the value given in Table 2 of 
Richtler (1990).  The near-integer offset suggests a typographical error
in Richtler's
Table; this was confirmed by comparison to the photometry of Isserstedt (1975),
which shows V = 12.8 for this star, rather than 13.8 as reported by 
Richtler (1990).  With this correction, Sk $-$66$\arcdeg$82 agreed perfectly
with our calibration relation.  Our reduced data are shown in Table \ref{photab};
column 1 contains an identification number assigned by ALLFRAME, columns 2 and 3
give the positional offsets in arcseconds north and west of the corner of the
field, and the photometric quantities $y$, $b-y$, and $m_1$, with associated
standard errors, fill the remainder of the table.

\begin{figure*}[t]
\centerline{\hbox{\epsfig{figure=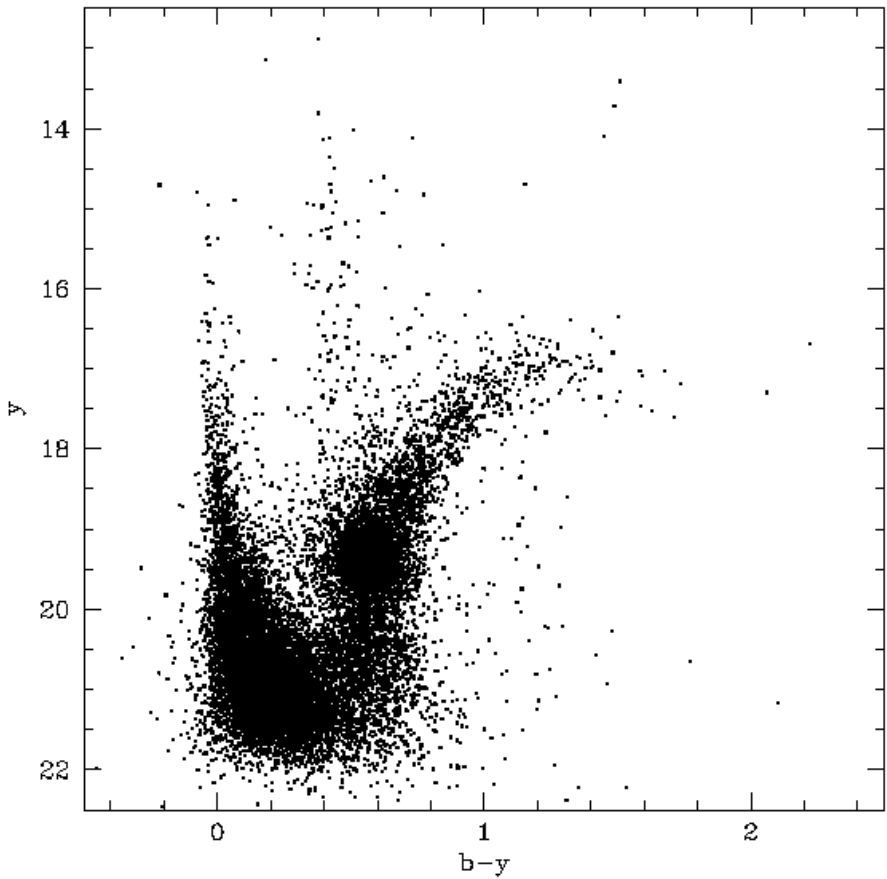,height=5.0in}}}
\caption{Calibrated ($b-y$, $y$) CMD for 16,958 stars in 
the Disk 1 field.
\label{bycmd}}
\end{figure*}

The calibrated ($b-y$, $y$) color-magnitude diagram is shown in 
Figure \ref{bycmd}.  The CMD shows 16,958 stars in the Disk 1 field, with
high completeness down to at least a magnitude below the level of the 
clump.  The highest density of points lies at the red clump, with an
extension to fainter, redder magnitudes due to the RGB-bump 
(as discussed by, e.g., Thomas 1968; Iben 1968; Alves \& Sarajedini 1999).
The upper giant branch shows the familiar broadening due to the mixture of 
ages and metallicities in the LMC (e.g., Stryker 1984).  
Few, if any, luminous AGB stars or 
horizontal branch stars are present in Fig. \ref{bycmd}.  The upper 
main-sequence is prominent for $y$ $\gtrsim$ 17.5, indicating the
presence of stars at least as young as 150 million years in this field.
The sparse, vertical plume of stars at $b-y$ $\approx$ 0.4--0.5 contains
primarily Galactic foreground dwarfs, with a small contribution from
blue loop supergiants in the LMC.

\begin{figure*}[t]
\centerline{\hbox{\epsfig{figure=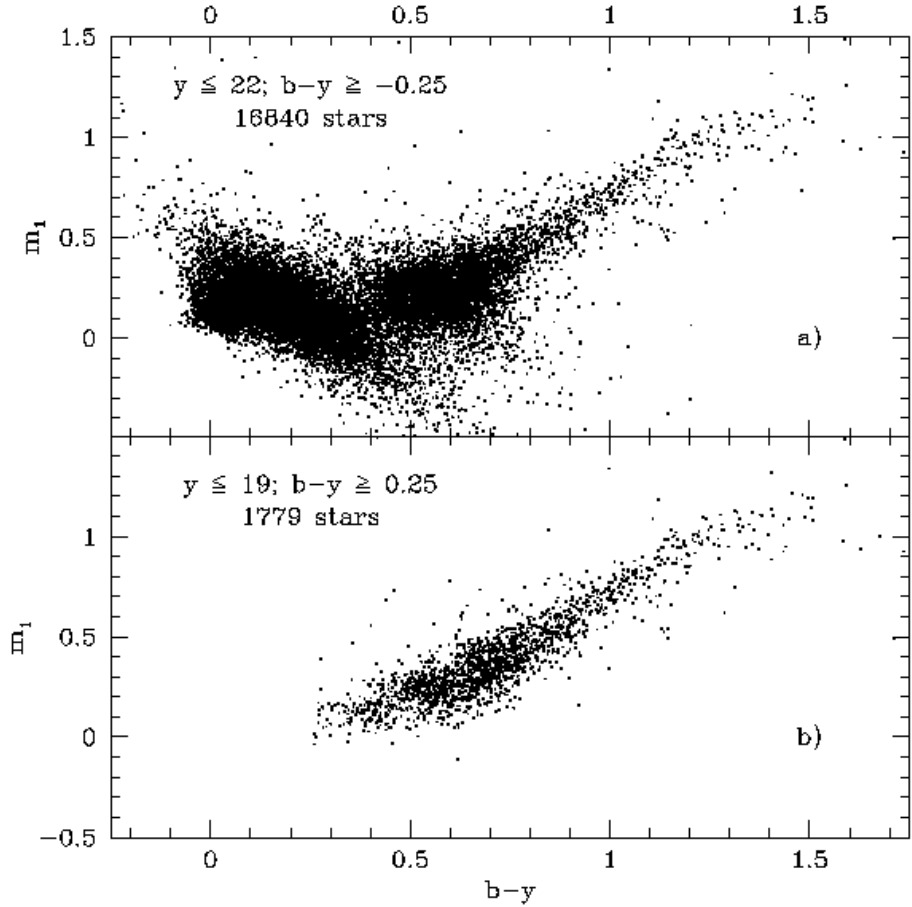,height=5.0in}}}
\caption{{\it a)} The color-metallicity plot (CMP) for 
the Disk 1 field.  {\it b)} The CMP, restricted to stars on the 
upper giant branch, and AGB.  These are the stars for which the 
metallicity calibration of Hilker (2000) is valid.
\label{cmp1}}
\end{figure*}

The ($b-y$, $m_1$) plot is shown in Figure \ref{cmp1}.  This diagram will
be referred to as the ``color-metallicity plot'' (CMP) for the remainder
of the paper.  The CMP will be the primary tool we use to determine the
the chemical abundances of red giant stars.  
Figure \ref{cmp1}a shows the CMP for our main-sequence and red giant
stars.  Note the narrow width of the giant branch as 
compared to Figure \ref{bycmd}; as described above,
age effects are not expected to contribute to the width of the RGB 
in the ($b-y$, $m_1$) plane.  Also note the large scatter
due to increasing errors for the faintest stars.  The error envelope
for the red clump stars has the potential to lead to overestimates of
the range of abundances contributing to the giant branch MDF.  
Because increasing photometric errors lead to biased estimates of
the range of abundances, we restrict our analysis to the upper giant
branch alone; these stars are shown in Figure \ref{cmp1}b.  For these
stars, the calibration of Hilker (2000) is expected to yield
accurate abundance estimates (0.5 $\leq$ ($b-y$) $\leq$ 1.1).

\subsection{Interstellar Reddening Corrections \label{ircsec}}

Because the determination of metallicities via the CMP relies
on a relation between a pair of stellar colors which differ in
their sensitivity to interstellar reddening, our results are 
sensitive both to the total amount of dust along the line of sight and
to the form of the reddening curve.  Because the $m_1$ index is
formed from three magnitudes, two of which are blue, its response
to interstellar attenuation is especially dependent on the shape
of the reddening curve \citep{cra76}.  

The most commonly cited values for the Str\"{o}mgren color excesses
are based on Crawford \& Barnes (1970): 
\eby\ = 0.74\, \ebv; \evby\ = $-$0.25\, \ebv.  These values
were based on the Whitford (1958) extinction curve.  Although this
curve is steeply rising into the violet spectral region, the 
relatively 
long wavelength baseline between the $b$ and $y$ filters drives
\evby\ to negative values.  We thus anticipate a need to ``debluen'', 
rather than deredden, our derived $m_1$ indices.

More modern reddening laws (e.g., Cardelli, Clayton \& Mathis 1989) 
show substantially
different selective absorption at the wavelength of the $b$ filter
than the Whitford (1958) curve; the adoption of a new reddening law
requires a recalculation of the Str\"{o}mgren color excesses.
However, the recent literature shows disagreement as to the effect
of interstellar reddening on the Str\"{o}mgren indices.  Schlegel et al. (1998)
and Fitzpatrick (1999), for example, approximately recover the Crawford \&
Mandwewala (1976) 
results: \eby\ = (0.75 $\pm$ 0.02)\, \ebv; 
\evby\ = ($-$0.25 $\pm$ 0.02)\, \ebv.  On the other hand, the model
atmosphere grids of \cite{kur99} show \eby\ = 0.69\, \ebv, an 8\% decrease,
and \evby\ = $-$0.12\, \ebv, a 50\% increase.

Because it was not to clear to us which of these determinations, if any,
are correct, we redetermined the Str\"{o}mgren system color excesses.
We convolved theoretical model stellar atmospheres appropriate to slightly
metal-poor ([M/H] = $-$0.5) K-type giants \citep{kur99} with the throughput curve of
the S2K CCD and the transmission efficiency curves of the CTIO Str\"{o}mgren
filter set.  When the resulting spectral energy distribution was convolved
with a Cardelli et al. (1989) extinction curve, 
assuming R$_V$ = 3.1, the results tabulated
by Kurucz (1999) were approximately recovered.  Our determinations of the 
Str\"{o}mgren color excesses are:

\begin{eqnarray}
\label{redlaw}
E_{b-y} = 0.70\, E_{B-V} \\
\label{rl2}
E_{m1} = -0.10\, E_{B-V}.
\end{eqnarray}

On the observational side, Delgado et al. (1997) used
the Str\"{o}mgren filter system to observe four highly
reddened open clusters.  In their reductions, they
derived \eby\ = (0.72 $\pm$0.04)\, \ebv, and
\evby\ = ($-$0.14 $\pm$0.04)\, \ebv.  The scatter
is consistent with {\it all} published determinations of
\eby; the observed \evby\ is 1$\sigma$ from our theoretically
calculated value, but $>$2.5$\sigma$ from the Schlegel et al. (1998)
or Fitzpatrick (1999) values.
The difference in spectral energy distribution between the
relatively blue stars observed by Delgado et al. (1997) and the 
cool model atmospheres \citep{kur99} 
is clearly not the source of the discrepancy between our values
for the Str\"{o}mgren color excesses and those of Schlegel et al. (1998) 
or Fitzpatrick (1999).

It is uncertain which set of reddening
values is more appropriate to observations of the LMC; we adopt
our calculated reddening law (Equations \ref{redlaw} and \ref{rl2}).
For an assumed reddening \ebv\ = 0.10 mag, the derived
[M/H]\footnote{We refer to the abundance
estimates from Str\"{o}mgren photometry as [M/H] rather than [Fe/H],
because the sensitivity of $m_1$ is to a combination of Fe and CN lines
\citep{bel78}.} for a given ($b-y$, \vbby) pair will be $+$0.08
dex higher if the Fitzpatrick (1999) values are used
for \eby\ and \evby.

Isofers in the CMP lie nearly perpendicular
to the reddening vector, so the derived abundances are highly
sensitive to errors in the adopted reddening.  For example,
a given ($b-y$, $m_1$) pair may yield an abundance [M/H] =
$-$0.33 if no reddening is assumed, but if dereddened by
\ebv\ = 0.10 mag, the derived abundance will be solar
([M/H] = 0).

We now turn to the determination of \ebv\ itself.  Schlegel et al. (1998) give
$\langle$\ebv$\rangle$ = 0.03 for this region of the LMC.  This is in
good agreement with the mean value for cool stars determined by Zaritsky (1999)
for a similar area on the opposite side of the LMC bar.  Given the 
importance of an accurate determination of \ebv, we used the bright, blue 
stars in Figure \ref{bycmd} to make an independent reddening determination
for the Disk 1 field.  We have constructed a reddening-free index for the
Str\"{o}mgren colors based on the relation E$_{v-b}$ = 0.82\, \eby.  We 
performed a linear fit to Kurucz model atmospheres for main-sequence
stars hotter than ($b-y$)$_0$ $\leq$ 0.10, and derived 

\begin{equation}
(v-b)_0 = 1.78\, (b-y)_0 + 0.173.
\end{equation}

If we define the reddening-free index for hot stars

\begin{equation}
Q_{vby} = (v-b) - 0.82\, (b-y),
\end{equation}

we then obtain the relation

\begin{equation}
\label{eqred}
(b-y)_0 = 1.04\, Q_{vby} - 0.173.
\end{equation}

We applied Equation \ref{eqred} to a sample of 412 stars bluer than
$b-y$ = 0.12 and brighter than $y$ = 19 to determine the mean reddening
in the Disk 1 field.  We find $\langle$\ebv$\rangle$ = 0.03 $\pm$0.01, 
with an rms scatter about the mean of $\pm$0.06 mag.
This is a very narrow distribution, given the $\approx$4\% accuracy of
our photometry.  Given the low mean value of \ebv\ and the almost 
unresolved scatter about that mean, we are confident that our 
abundance estimates will not be strongly affected by reddening issues.
Our adopted mean reddening values in the Str\"{o}mgren colors are
\eby\ = 0.021, \evby\ = $-$0.003.

\subsection{Calibrating the Color-Metallicity Relation \label{cmrsec}}

\begin{figure*}[t]
\centerline{\hbox{\epsfig{figure=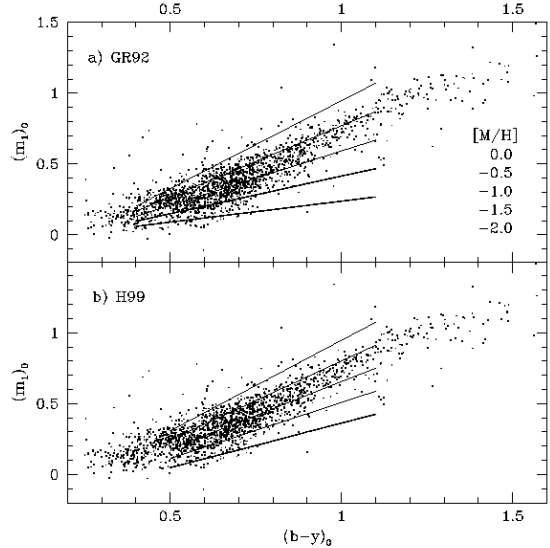,height=3.0in}}}
\caption{Dereddened CMP for the Disk 1 field.  A reddening
of $\langle$\ebv$\rangle$ = 0.03 has been adopted.  Isofers from two
different calibrations are shown: {\it a)} Grebel \& Richtler (1992), valid for
0.4 $\leq$ ($b-y$)$_0$ $\leq$ 1.1; {\it b)} Hilker (2000), valid for
0.5 $<$ ($b-y$)$_0$ $<$ 1.1. 
\label{cmp2}}
\end{figure*}

Our technique for using the CMP to derive abundance estimates is based
on the observation of Bond (1980) that the slope of the RGB in the 
CMP is a nearly linear function of [M/H].
This relation was calibrated
for metal-rich giants and supergiants by Grebel \& Richtler (1992); the calibration
was very recently extended to low metallicity by Hilker (2000).  In 
Figure \ref{cmp2} we show our dereddened CMP, with isofers from Grebel \& Richtler (1992) 
overplotted in the top panel, and from Hilker (2000) in the bottom panel.
Isofers are plotted at 0.5 dex intervals between 0.0 $\geq$ [M/H] $\geq$
$-$2.0, in both panels.  Hilker (2000) quotes a scatter around his
mean CMR of $\pm$0.16 dex; this is comparable to the scatter in the
Grebel \& Richtler (1992) calibration, of $\pm$0.22 dex.  These values limit the
precision with which the abundance of an individual red giant can be 
measured using the Str\"{o}mgren CMP.

The two calibrations agree for solar metallicity, but the 
Hilker (2000) relation shows a steeper slope for low metallicities.  
We performed a simple check on the color-metallicity relation (CMR) using
stellar evolutionary tracks from the Padua group \citep{gir00}, convolved
with model atmosphere grids \citep{kur99} and the CTIO Str\"{o}mgren
filter response functions, for stars of 1 and 2 M$_{\sun}$.  We found that
both CMRs match the theoretical tracks for [M/H] $\geq$ $-$0.5, but there
is a progressive steepening of the RGB slope as the abundance decreases,
which agrees with the steeper CMR of Hilker (2000).
The theoretical tracks also start to show deviations from linearity at
the blue end of the Grebel \& Richtler (1992) calibration for low metallicity, justifying
Hilker's (2000) restriction to ($b-y$)$_0$ $>$ 0.5.  However, for [M/H]
$\gtrsim$ $-$0.5, the RGB remains linear at least to the ($b-y$)$_0$ = 0.4
limit \citep{gre92}. 
At the red end of the calibration, 
the theoretical tracks predict that the RGB deviates strongly from
linearity for the metal-rich stars, as the $b$-band becomes progressively
more swamped by molecular opacity.  In the models, this deviation from the
linear CMR is seen for solar-metallicity stars cooler than ($b-y$)$_0$ = 0.9.
Such cool, high-abundance stars could possibly account for the group
of several stars at ($b-y$)$_0$ $\approx$ 1.1, ($m_1$)$_0$ $\approx$ 0.5;
however the high-metallicity RGB does not appear sufficiently well-populated
blueward of ($b-y$)$_0$ = 0.8 to produce these stars.
In general, Figure \ref{cmp2} shows only a mild flattening of the
RGB beyond ($b-y$)$_0$ = 1.1, if any, due to the subsolar metallicity of the
LMC field.

An unfortunate property of the Str\"{o}mgren CMR
is the convergence of isofers in the
region of the red clump.  Where the isofers converge, small
errors in photometry and shifts due to differential reddening
will lead to huge errors in derived abundance.  This effect
is stronger in the Grebel \& Richtler (1992) CMR, simply because their
isofers converge more rapidly at high temperatures. At the 
level of the red clump, a typical 3\% error in $b$ can 
lead to an abundance error of $\pm$0.84 dex.  As the 
isofers diverge towards cooler temperatures, this becomes
a less significant source of error.  However, most of the 
RGB stars lie in the red clump region (see Figure \ref{bycmd});
the convergence of isofers is the strongest
limitation of the Str\"{o}mgren CMP for abundance determinations
of large samples of stars.

For the remainder of this paper, we adopt the Hilker (2000) color-metallicity
relation, because it has been more extensively compared to metal-poor star
clusters and  provides a somewhat better match to theoretical models for 
low abundance.
The difference between the GR92 and
H99 color-metallicity relations is a function of temperature and abundance.
For example, at ($b-y$)$_0$ = 0.5, ($m_1$)$_0$ = 0.25, the difference
$\Delta$[M/H]$_{GR92-H99}$ = $-$0.20.  In contrast, at 
($b-y$)$_0$ = 1.0, ($m_1$)$_0$ = 0.60, $\Delta$[M/H]$_{GR92-H99}$ = $+$0.20.
For a ``typical'' star in our Disk 1 field, at ($b-y$)$_0$ $\approx$ 0.65
and ($m_1$)$_0$ $\approx$ 0.4, the difference $\Delta$[M/H]$_{GR92-H99}$ 
$\lesssim$ $\pm$0.05, smaller than the typical measurement error of 
$\approx$ $\pm$ 0.2 dex.

\subsection{The Photometric Metallicity Distribution Function \label{pmsec}}

\begin{figure*}[t]
\centerline{\hbox{\epsfig{figure=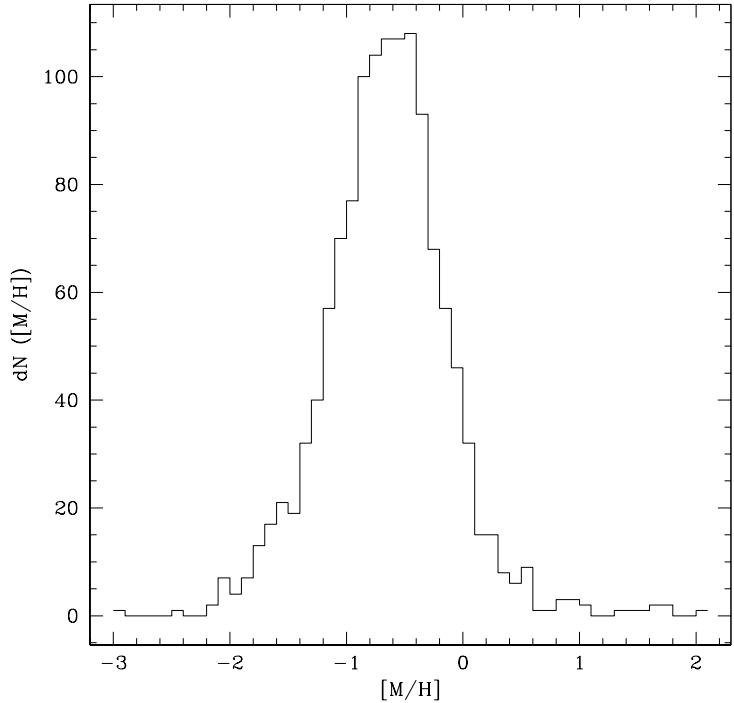,height=4.0in}}}
\caption{The photometric metallicity distribution
function for the Disk 1 field, using the Hilker (2000) color-metallicity
relation.  The mean [M/H] = $-$0.649 $\pm$0.005 for the 1261 stars in
the histogram.
\label{h99fig}}
\end{figure*}

The binned MDF of 1261 red giants in the Disk 1 field is shown in 
Figure \ref{h99fig}.  Only stars brighter than $y$ = 19 
and redder than $b-y$ = 0.50 have been
included in the determination.  The MDF is characterized by a 
broad peak between [M/H] = $-$0.7 and $-$0.4, with a steep fall-off
towards higher values and a shallower tail to low abundances.
In this respect, we have recaptured the basic features of the spectroscopic
MDF.
The most probable abundance is [M/H] = $-$0.55 $\pm$0.10, while
the mean is [M/H] =  $-$0.649 $\pm$ 0.005.  The median [M/H] = 
$-$0.635, while the 1st and 3rd quartile values are $-$0.965 and
$-$0.342, respectively.  While just 6\% of the stars are more 
metal-poor than [M/H] = $-$1.5, 8\% of the stars show abundance
above solar.  Contamination by Galactic foreground stars
may account for most of the super-metal-rich stars\footnote{The 
effect of a photometric blend between an upper RGB star and
a main-sequence or red clump star shifts its metallicity to
lower, not higher abundances.} (see \S \ref{fgsec} below). 

The 1$\sigma$ width of the best-fitting Gaussian to the MDF
is 0.44 dex.  However, a single Gaussian leaves a large
residual below [M/H] = $-$1.5 unaccounted for; a much better fit was
achieved using two Gaussians, a metal-rich and a metal-poor component.
A bimodal MDF is reasonable, considering determinations based on 
star clusters \citep{ols91}.
The metal-rich component, containing 92\% of the stars, has a mean
[M/H] = $-$0.625 $\pm$0.011 with a 1$\sigma$ width of 0.42 dex.  The
metal-poor component accounts for the remaining 8\% of the stars, with
mean [M/H] = $-$1.64 $\pm$0.09 and $\sigma$ = 0.29 dex.  The width
of the metal-rich component is much greater than that expected from
random errors; that of the metal-poor component is less well-resolved.

If we had adopted the Grebel \& Richtler CMR, the mean [M/H] would have
been shifted downward to $-$0.677 $\pm$0.006, with a more sharply
peaked distribution around [M/H] = $-$0.35.  The locations
of the quartile points, were changed by $\approx$0.03 dex, and 
the fraction of stars above [M/H] = 0 or below [M/H] = $-$1.5 were
unchanged.

\subsubsection{The Effect of CN-Anomalous Stars \label{cnsec}}

A drawback of the Str\"{o}mgren $m_1$ index is its sensitivity
to variations in CN abundance \citep{bel78}.  CN can even dominate [Fe/H]
in determining the $m_1$ of globular cluster giants \citep{ant95}.
In metal-poor globulars, a wide variation in CN strengths is often
seen for nearly constant [Fe/H]; the CN-strength distribution 
function is often bimodal \citep{kra94,ant95}.  
Such a bimodal
distribution of CN strengths could be invoked to explain the
possible secondary component of our photometric Disk 1 MDF, with
the majority of the stars populating the CN-strong peak.  Without
high-resolution spectra, it is impossible to evaluate this 
hypothesis conclusively. 

Because of the multiple dredge-up
episodes that RGB and AGB stars undergo, the CN abundances in 
planetary nebulae have not fully constrained the patterns of CN variation
in the still-evolving RGB stars of the LMC field \citep{dop97}.
We note that the stars in our field star sample, less massive than
4 M$_{\sun}$ and less luminous than the tip of the red giant branch,
have experienced neither 2nd \citep{ren81} nor 3rd \citep{gir00}
dredge-up events.  However, their CN abundances may have been 
altered by the 1st dredge-up episode, and/or subsequent ``extra'' 
mixing \citep{cha94}.

While CN variations may contribute to
the width of Figure \ref{h99fig}, we doubt that this effect is
capable of shifting stars from the primary peak at [M/H] $\approx$
$-$0.5 into the secondary peak at [M/H] $\approx$ $-$1.6.
Figure 7 of Kraft's (1994) review indicates that the separation 
between CN-strong and CN-weak peaks in a given globular cluster are
less than $\approx$0.5 dex, rather than the $\approx$1 dex separation
between the two peaks in our Figure \ref{h99fig}.
Additionally, mildly metal-poor stars ([Fe/H] $\gtrsim$ $-$1) in 
the Galaxy show far less variation in CN strengh than do the more 
metal-deficient stars \citep{kra94}.  Studies of old Galactic 
open clusters have found very little evidence for variations of
CN strengths within a given cluster \citep{jan84a,jan84b}.  Most
of the LMC disk stars are closer in age and abundance to the open
clusters than the globulars, but we cannot yet with certainty rule
out CN anomalies as a cause of structure in the MDF.  In section
\ref{dsec} below, we show that there is evidence that the metal-poor
tail of Figure \ref{h99fig} may in fact be stars with 
[Fe/H] $\approx$ $-$1 that are relatively weak in CN.

\subsubsection{Differential Reddening \label{drsec}}

Differential reddening is expected to be very small in our Disk 1
field, based on 2MASS color-color plots \citep{nik00}.  This confirms
expectations based on inspection of the H {\small I} aperture synthesis
maps of Kim et al. (1998), which show our field to be relatively
lacking in small-scale structure in the neutral ISM.
While differential reddening is expected to be a minor effect for the
Disk 1 field in general,
it may have a strong effect on the measurements of individual
stars.  Many LMC red giants suffer essentially zero reddening \citep{zar99};
if they are erroneously dereddened by \ebv\ = 0.03 mag, the resulting
[M/H] measurement will be in error by $+$0.12 dex, on average.  Similarly,
the distribution of observed reddening values shows a tail to high 
(A$_V$ $\gtrsim$ 1) values; a star obscured to such a degree would have 
its abundance underestimated by $\sim$1.3 dex, i.e., a factor of 20.
Clearly the use of the color-metallicity plot should not be relied
upon for accurate 
abundance measurements of individual stars unless independent reddening
information is also available.
However, insofar as most stars in the
field are lightly reddened by an average \ebv, the Str\"{o}mgren CMR
will yield an accurate value of the mean abundance.  


\subsubsection{Subtracting the Foreground Dwarfs \label{fgsec}}

Because the Str\"{o}mgren CMP contains no leverage on surface gravity,
our photometric MDF is contaminated by late-type, main-sequence
stars in the Galactic foreground.  The dominant foreground population 
is the vertical plume of old stars at the main-sequence turnoff corresponding
to the age of the (thick) disk (9--11 Gyr: e.g, Morell et al. 1992).
These stars can be seen in Figure \ref{bycmd}
at $b-y$ $\approx$ 0.4, the color expected for G-type dwarfs \citep{cra70}.
These stars mainly lie to the blue of the area we analyze,
and thus only weakly contaminate our photometric MDF.

Cooler foreground dwarfs lie in a tenuous veil across the CMD redward 
of $b-y$ = 0.4, out to the bottom of the main-sequence somewhere in the
vicinity of $b-y$ = 1.3.  Their numbers increase dramatically towards
fainter magnitudes \citep{all73,bah80}.  In order to estimate the
level of contamination of the MDF by late-type dwarfs, we compared
starcounts in Figure \ref{bycmd} to the predictions of 
Ratnatunga \& Bahcall (1985).
Their model predicts that 55, or $\approx$4.4\% of the stars that went
into our determination of the MDF, are foreground K--M dwarfs. 

The Ratnatunga \& Bahcall (1985) starcounts are 
based on the model of Bahcall \& Soneira (1980), which 
is highly smoothed relative to the real, clumpy distribution of stars
on the sky; hence variations of a factor of a few from the model 
predictions are not uncommon.  We can scale Ratnatunga \& Bahcall's (1985)
prediction to the observed foreground stellar distribution by
examining the stars brighter than $y$ = 16.5 and redder than $b-y$ = 
0.1, where the contribution of bona fide LMC giants is expected
to be negligible.  For $y$ $\leq$ 15, where the model predicts 12.5 foreground
stars, we count 20, a 60\% excess that is only expected to occur by chance
in 1.3\% of trials.  If we include stars down to $y$ $\leq$ 16.5, we count
62 stars where 39.5 are predicted, also a 60\% excess.   The probability
P(62;39.5) = 0.02\%, from Poisson statistics; therefore we conclude there
is a highly significant excess of foreground stars above the Ratnatunga \& Bahcall (1985)
prediction.  We would then expect nearly 90 contaminators in our photometric
MDF, some 7\% of the sample.

To subtract the foreground contaminators from the MDF, it is necessary
to assume a statistical form for the MDF of the Galactic 
disk\footnote{Halo stars are a negligible fraction of the total along
this line of sight (Bahcall \& Soneira 1980).}.  Wyse \& Gilmore (1995) give the combined MDF
of the
thin and thick disk, showing that it peaks near [M/H] = $-$0.2, with a range of 
$-$1.4 $\lesssim$ [M/H] $\lesssim$ $+$0.2.   Convolving the intrinsic
MDF of the thin$+$thick disk from Table 3 of
Wyse \& Gilmore (1995) with the $\pm$0.16 scatter about
the color-metallicity relation, we expect to observe $\approx$15
foreground stars more metal-rich than [M/H] = 0; most of the genuinely
super metal-rich stars in our MDF (as opposed to normal stars that appear
super metal-rich due to photometric error) are likely to be foreground
dwarfs.

The mean [M/H] of 
a sample of stars drawn from the Wyse \& Gilmore (1995) MDF is $\approx$ $-$0.33;
subtracting these stars from Figure \ref{h99fig} shifts the mean 
[M/H] to $-$0.675 (a shift of $-$0.026 dex), does not affect the
1st quartile point, and shifts the 3rd quartile point downward by
$-$0.07 dex, to $-$0.41.

\section{Discussion \label{dsec}}

\subsection{The Metallicity Distribution Function of LMC Red Giants 
\label{mdfsec}}

The problem with the schemes for correcting the uncertainties in our 
observed spectroscopic MDF is that they themselves rest on uncertain assumptions
and circular reasoning.  In fact, the adoption of a specific prescription
to correct for the known systematics of our spectroscopic MDF is likely
to introduce more problems than it solves.  As an example, our corrections
for the fading of the red clump and for the [Ca/Fe] ratio assumed a 
dispersionless age-metallicity relation; such a thing does not exist,
according to deep color-magnitude diagram data \citep{hol99}.  Therefore
we consider only the 
uncorrected metallicity distribution function here.  However, we have
shown in section \ref{corsec} that {\it the main sources of systematic
error do not change the basic form of the MDF}.  Uncertainties in the
relative height and width of the $-$0.55 dex peak, and in the value for
the high-metallicity cutoff surely remain, at the $\pm$0.1--0.2 dex level.
Nevertheless, we have identified the basic features of the MDF with enough
certainty to discuss its form and implications.

\begin{figure*}[t]
\centerline{\hbox{\epsfig{figure=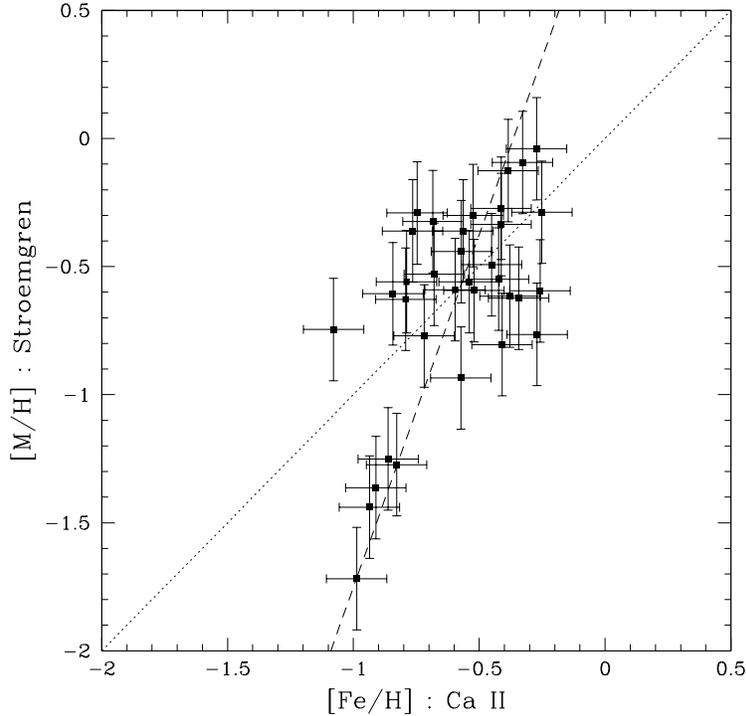,height=4.0in}}}
\caption{Comparied of \caii\ triplet-derived
abundances to Str\"{o}mgren-derived abundances for 34 stars measured
by both techniques.  {\it Dotted line:} The line of equivalence;
{\it dashed line:} The best-fit linear relation [Fe/H] = 
0.363[M/H] $-$ 0.364.  The scatter about the linear fit 
$\sigma$ = 0.18 dex.
\label{starfig}}
\end{figure*}

First we compare the MDFs derived from photometry and spectroscopy.
The advantage of photometric abundance determinations is the ease with 
which large numbers of stars can be measured.  It is of crucial importance
to be able place such measurements on a common scale with spectroscopic
studies.   Figure \ref{starfig} compares the abundances derived from the 
\caii\ triplet vs.\ those derived from Str\"{o}mgren photometry for
the 34 stars for which both types of data were available and which
fell within the color range of the Str\"{o}mgren metallicity 
calibration \citep{hil99}.  The line of equivalence is shown as a dotted
line, along with a dashed line representing a linear, least-squares
fit to the data.  

In order to gain advantage from the large numbers of stars in
photometric samples, the random errors of measurement must exceed
any serious systematic errors, thereby allowing the errors in 
statistical quantities to be reduced by observing many targets.
The total random error in the Str\"{o}mgren measurement is dominated
by photometric errors (Poisson noise, aperture correction,
transformation error, and crowding error); conservatively (i.e.,
possibly underestimating the error), these contribute $\approx$
$\pm$0.25 dex to the measurement of an individual star.  
Variable reddening and scatter in the calibration relation contribute
an additional $\pm$0.12 and $\pm$0.16 dex of scatter, respectively,
for a total random error of $\sigma$ $\gtrsim$ 0.32 dex.  The two
main sources of systematic error in the Str\"{o}mgren MDF are
variations in the reddening law and variations in CN on the giant
branch; the first of these contributes a maximum error of 0.05
dex, because of the low mean reddening in this field.  CN-anomalies
are likely to contribute less than $\pm$0.1 dex to the error budget
\citep{jan84b}.  Thus, while systematics cannot be ignored, they
are a minority contributor to the total error, and thus the
multiplexing advantage of CCD photometry is retained.  It is 
important to note that in very highly reddened fields, for old
and metal-poor stars (in which CN-anomalies are large and common),
the balance of this equation may tip the other way.

We define the offset between the two abundance measures to be
\dmh\ $\equiv$ [Fe/H]$_{Ca}$ $-$ [M/H]$_{vby}$.  The mean value 
of \dmh\ is $-$0.03 dex, with a 1-sigma scatter of $\pm$0.31 dex
about the line of equivalence.  This is well within the range 
expected given the errors in both techniques.  However, the offsets
are strongly correlated with metallicity.  We find 

\begin{equation}
\label{sceqn}
\Delta \left[\frac{M}{H}\right] = 0.637\, [M/H] + 0.365,
\end{equation}

\noindent with an rms scatter about the line of just $\pm$0.18 dex.
This formula can be used to correct Str\"{o}mgren-derived abundances
in order to place them on approximately the same scale as the 
higher-precision spectroscopic measurements:

\begin{equation}
\label{coreqn}
[Fe/H]_{Ca} = 0.363\, [M/H]_{vby} - 0.364.
\end{equation}

\noindent  This is the dashed line shown in Figure \ref{starfig}.
The residuals to that fit are uncorrelated
with abundance.   The reason for the good statistical agreement
between the photometric and spectroscopic MDFs is that the 
large majority of stars in the sample lie near [Fe/H] = $-$0.57,
where the difference between the two methods vanishes.  
Although 
the scatter at the high-metallicity end of Figure \ref{starfig}
somewhat masks the correlation, we find a linear correlation 
coefficient between \dmh\ and [M/H] of $r = 0.81$.  The scatter
at high abundance is due to the systematic uncertainty in the
calibration of the \caii\ triplet method above [Fe/H] $\approx$ $-$0.5.
Additional scatter could be introduced by differential reddening or by 
variations in CN strength at a given [Fe/H], but is not required
to reproduce the data.

The cause of the metal-dependent offset between Str\"{o}mgren
and \caii\ triplet abundance scales is unknown.  For the few 
globular clusters measured with both techniques, offsets are
present, but are much smaller than that inferred from Figure \ref{starfig}
(e.g., Richtler et al.\ 1999).  Because the globular clusters
are much older than the bulk of LMC field stars, age effects may be
suspected.  However, the Str\"{o}mgren color-metallicity plot is
predicted to be insensitive to age (\S \ref{phsec}).  And although the
\caii\ triplet is sensitive to surface gravity \citep{jor92}, we 
have shown in \S \ref{corsec} that age effects are likely to be
small except for the youngest and most metal-rich stars in our 
sample, not exceeding 0.1--0.2 dex at the most.  The direct
calibration of the \caii\ abundances to the globular cluster
abundance scale makes these values more reliable than the 
photometric indices, especially below [Fe/H] $\approx$ $-$0.5.

If age effects are discounted, then the source of the discrepancy
is likely to lie with abundance ratio anomalies; some process is
acting to decrease the (Fe$+$CN)/Ca ratio with decreasing [Fe/H]
among the LMC field stars.  Figure \ref{cafefig} shows that the
chemical evolution model of Pagel \& Tautvai\v{s}ien\.{e} (1998)
predicts an increasing contribution of calcium to the heavy element
fraction as the metallicity decreases.  However, two factors militate
against this being the sole contributor to \dmh : first, the effect
is nonlinear with [Fe/H], unlike \dmh ; and second, the effect is 
unlikely to exceed 0.1 dex over the metallicity range of interest.

While calcium abundances are unaffected by mixing processes
during stellar evolution, carbon and nitrogen abundances may be 
altered by dredge-up of the products of CNO-burning \citep{ren81}.
The stars in our spectroscopic sample have all experienced the
1st dredge-up episode, which occurs at the base of the red giant
branch and acts to deplete the surface abundance of $^{12}$C while
enhancing $^{14}$N.  None of 
the stars in our spectroscopic sample have been
affected by 2nd dredge-up (which acts similarly to 1st dredge-up),
or 3rd dredge-up (which strongly enhances $^{12}$C, the primary product
of helium-shell burning).   In addition to the dredge-up processes,
``extra'' mixing may take place in some or all RGB stars, above the
RGB-bump; all of our stars lie beyond this point.  
Both 1st dredge-up and extra mixing are predicted to be functions
of metallicity \citep{cha94,boo99}.   Thus it may be possible 
that metallicity-dependent mixing processes are acting to 
deplete CN more quickly for metal-poor stars than for metal-rich stars,
leading to increasingly more positive values of \dmh\ as [Fe/H]
decreases.  

Possible support for this can be gleaned from Figure
7 of Dopita et al.\ (1997), where the abundance ratios of carbon,
nitrogen, oxygen, and helium are plotted as functions of 
[$\alpha$/H] for planetary nebulae.  The figure shows that
[N/$\alpha$] drops precipitously with [$\alpha$/H]; and that
while [C/$\alpha$] rises during the same overall abundance range,
most of the increase is due to 3rd dredge up, which does not
apply to our stellar sample.  In star clusters, it is known
that CN abundance is usually correlated with the nitrogen
abundance, and anti-correlated with carbon
\citep[and references therein]{smi92}.
The observed drop in final nitrogen
abundance with decreasing metallicity suggests that mixing
is acting much less efficiently in low-metallicity stars; 
Smith (1992) showed how inefficient dredge-up of nitrogen
would lead to CN-weak stars.  The strong drop in [N/$\alpha$] 
for the Dopita et al.\ (1997) sample of PNe may indicate
that the relatively metal-weak field stars of the LMC are
generally CN-weak.  However, this result is somewhat 
at odds with predictions 
that mixing processes on the upper RGB are much more 
efficient at lower abundance than at higher abundance
(e.g., Boothroyd \& Sackmann 1999); this 
would tend to drive the CN-based abundance measures {\it higher}
than the Ca-based measures for metal-poor stars, contrary
to our observations.  To confirm or reject the interpretation
of Figure \ref{starfig} as the result of metallicity-dependent
mixing would require high-resolution spectroscopy for detailed
abundance analyses, as well as further Str\"{o}mgren photometry
of intermediate-age, metal-poor populations, preferably extending
to magnitudes well below the RGB-bump (V $\approx$ 19.4 in the
LMC field).

\begin{figure*}[t]
\centerline{\hbox{\epsfig{figure=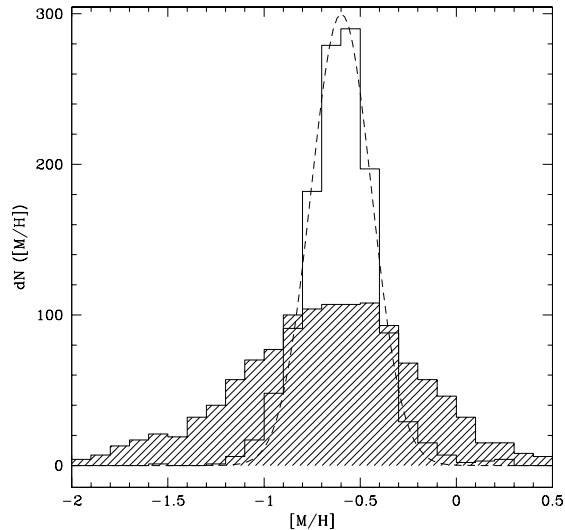,height=3.0in}}}
\caption{The photometric MDF (figure \ref{h99fig}),
 shown as the shaded histogram.  The histogram of photometric 
abundances, after correction based on the linear fit shown in 
Figure \ref{starfig}, is shown as the solid line; the best-fit
gaussian, with a mean of [M/H]$'$ = $-$0.60 and $\sigma$ = 0.20 dex,
is shown as a dashed line.
\label{cpmdf}}
\end{figure*}

Although its provenance remains uncertain, we 
can use Equation \ref{coreqn} to shift our photometric MDF 
(Figure \ref{h99fig}) onto the \caii\ triplet abundance scale, 
and to see whether the secondary peak in the photometric MDF
is an artifact of the metal-dependent offset between 
indices.
Unfortunately, none of the three stars with [Fe/H]$_{Ca}$ $<$ $-$1.1
fell into the region with Str\"{o}mgren
photometry, so Equation \ref{coreqn} is not strictly applicable 
to stars more metal-poor than [M/H] $\approx$ $-$1.7.  Still, that
range encompasses the peak of the putative metal-poor population discussed
in section \ref{pmsec}.  The result of the correction is shown in 
Figure \ref{cpmdf}: the uncorrected data are shown as the shaded 
histogram, while the corrected MDF is shown as the solid-outlined
histogram.  The effect of the correction is to sharply narrow the
distribution; a single Gaussian with mean [M/H]$'$ = $-$0.60 and
dispersion $\sigma$ = $\pm$0.20 is now a close fit to the data (dashed
curve in Figure \ref{cpmdf}).  There remains an excess of high-metallicity
stars, attributable to Galactic foreground.  The ``secondary peak'' discussed
in \S \ref{pmsec} is much less significant now; adding a second Gaussian
to the fit does not improve it significantly, although the amplitude of
8--10\% agrees with that derived earlier.  However, the mean of this 
secondary component is now [M/H]$'$ = $-$0.93, making it indistinguishable
from a slightly extended low-metallicity wing of the main population.
It is interesting
to note, however, that a second population at [Fe/H] $\approx$ $-$1 is in
better agreement with the spectroscopic MDF (Fig.\ \ref{ca2hist}) than
a population with [Fe/H] $\approx$ $-$1.6 would be.  The corrected
photometric MDF now seems to {\it underpredict} the number of metal-poor
stars, compared to the spectroscopic MDF; Str\"{o}mgren photometry of
stars below [Fe/H] = $-$1.1 is needed to address this deficiency.

Because of the uncertainties attending the form of our photometric
MDF, we prefer to explore the chemical evolution of the 
Disk 1 field using 
our spectroscopic MDF (Figure \ref{ca2hist}).
The shape of our MDF is similar to the MDF of F/G dwarfs
in the disk of the Milky Way (e.g., Wyse \& Gilmore 1995), shifted to lower
values by $\approx$ 0.3--0.4 dex.  In this respect, our Disk 1 field
closely resembles the disk of the Galaxy in its patterns of chemical
enrichment, although this enrichment has been retarded by a factor
of $\approx$2 in the smaller galaxy.  This pattern apparently extends
to the form of the age-metallicity relation, which in the LMC as for the 
Milky Way seems to have been a mildly increasing function for several
Gyr before an abrupt increase during the past 1--3 Gyr 
\citep{dac91,edv93,wys95,pag98}.   

The most metal-rich stars we found in the spectroscopic sample showed 
[Fe/H] $\approx$ $-$0.25; this is consistent with a cutoff in the photometric
sample if the decreased precision and contamination by foreground dwarfs
is considered.  Because we have measured red giant stars, our MDF does 
not extend to the youngest stars in the LMC, cutting off at $\sim$500 Myr.
If chemical enrichment has proceeded normally during the ensuing timespan,
then younger populations should be more metal rich than this limit.  This
behavior is in fact observed: Hill et al. (1995) find [Fe/H] = $-$0.24 $\pm$0.1
for a sample of F/G supergiants, and Jasniewicz \& Th\'{e}venin (1994) note similar values
for 10--40 Myr-old star clusters.  The most metal-rich stars in the LMC
are about twice as enriched as the intermediate-age stars measured here, 
a fact which must be taken into account if these results are to be 
correctly applied to synthetic CMD analyses.

\subsection{Spatial Variations in Stellar Populations \label{varsec}}

Recent WFPC2 \citep{geh98,hol99} and ground-based \citep{wes98} studies
have suggested that most areas of the LMC disk share a common \sfh.
This is broadly consistent with a shallow or nonexistent radial
metallicity gradient \citep{ols91}.  These studies appear to conflict
with the earlier results of Bertelli et al. (1992) and Vallenari 
et al. (1996), who found evidence from wider-field, but shallower,
CMDs for a spatially varying \sfh.  Most of the differences may
be attributable to variations in adopted age-metallicity relation.
There is also increasingly strong
evidence that the form of the MDF varies spatially \citep{bic99}.
Those authors examine the spatio-temporo-chemical distributions of
star clusters and find that although the radially distant clusters
are more metal-poor than the inner clusters, they are also older.  For
clusters of a given age, Bica et al. (1999) do not find evidence for any
metallicity gradient.

This fact has been interpreted as evidence for two distinct populations
of star clusters (e.g, Kontizas et al. 1990).  In this picture, there is an 
inner disk, $\approx$8$\arcdeg$ across, embedded within a thicker 
disk $\approx$ 13$\arcdeg$ $\times$ 10$\arcdeg$ in size.  The data of 
Bica et al. (1999) can be placed within this framework, indicating that the
LMC has a spatially compact intermediate-age cluster system and a more
diffuse, older system (also see Olszewski 1999).  

\begin{figure*}[p]
\centerline{\hbox{\epsfig{figure=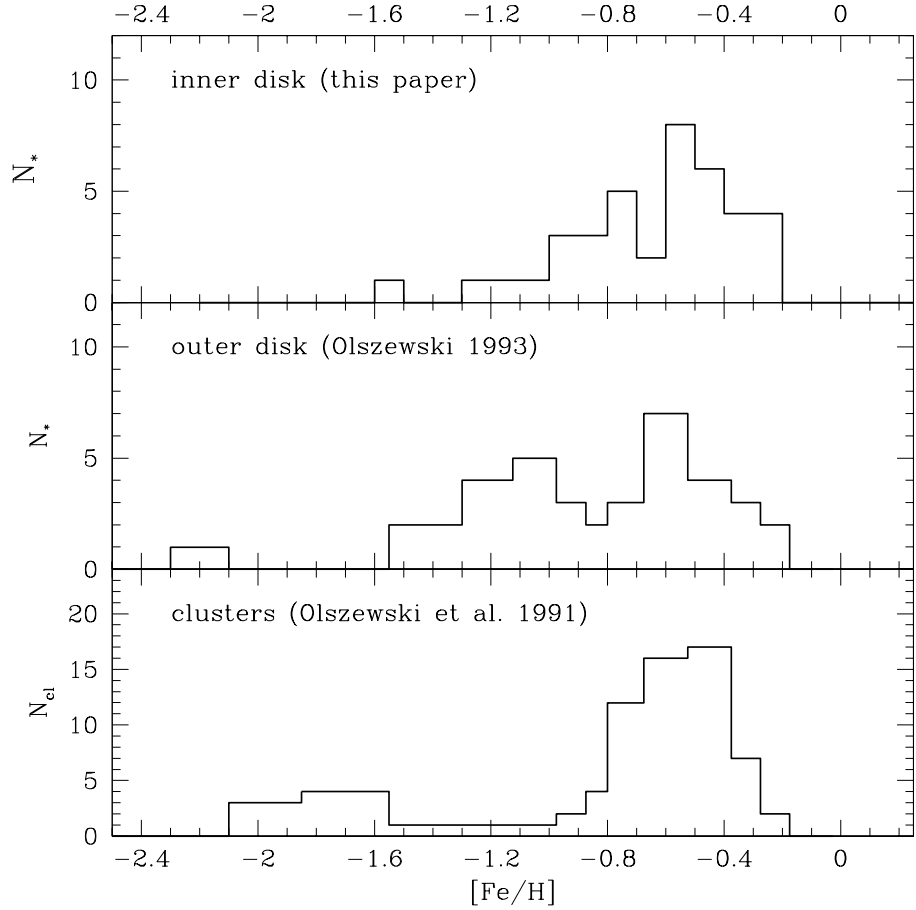,height=5.0in}}}
\caption{Our spectroscopic MDF (top panel), 
compared with that of 36 outer disk field stars from Olszewski 
(1993; middle panel), and 70 LMC star clusters from Olszewski 
et al. (1991; bottom panel).  The older data have been corrected
to the Carretta \& Gratton (1997) abundance scale as described
in the text.  The fraction of stars below [Fe/H] $\approx$ $-$1
is 3 times higher in the outer disk field than in the inner disk
field. The field stars of the inner disk lack the extremely 
metal-poor component represented by the ancient globular clusters,
but seem similar to the intermediate-age clusters.
\label{o9193fig}}
\end{figure*}

This model for the LMC structure is reminiscent of the ``Baade sheet'' 
concept for stellar populations, in which all galaxies support an
extended distribution of ancient stars \citep{hod89}.  Do the abundance
distributions of field stars support this pattern? Olszewski (1993) presented
spectroscopic data, obtained for the \caii\ triplet, for 36 red giants 
in a far northern field of the LMC.  Their field was in close proximity
to the globular cluster NGC 2257, 8$\fdg$5 northeast of the optical
center of the bar (i.e., at 5 times larger radius than our Disk 1 field).

We compare the abundance histogram of Olszewski (1993) to our spectroscopic
MDF in Figure \ref{o9193fig}.  Because their results were calibrated to 
the Zinn \& West (1984) abundance scale, we have corrected their histogram bins
to the high-dispersion, Carretta \& Gratton (1997) scale.  
This has been achieved using the relations in 
Rutledge et al. (1997a), for [Fe/H]$_{ZW}$ $\leq$ $-$0.5, and by taking a constant
offset between the two scales for higher abundances.  

To within the 
calibration uncertainties, the two abundance distributions are seen to 
agree well, above [Fe/H] $\approx$ $-$0.8.  Below that level, the difference
is striking.  The second maximum found  at [Fe/H] = $-$1.2 by Olszewski (1993)
is reduced to a mere tail in our sample.  While one-third of the outer disk
sample lies below [Fe/H] = $-$1 \citep{ols93}, only 10\% of the inner disk sample is
so metal-deficient.  There are two possible explanations for this.  One
is that selection effects have either biased us against detection of 
metal-poor stars, or that Olszewski (1993) were biased in favor of metal-poor 
stars.  Because we chose our target stars from a random sample of colors, it is 
unlikely that we would have missed such a large fraction of metal-poor stars.
Olszewski (1993) included stars well to the blue of the RGB ridgeline in their
sample, in an effort to obtain a complete census of metal-poor stars, and so
it is quite possible to believe that they more efficiently detected
metal-poor stars than we did.  However, we examined a color-magnitude
diagram of the NGC 2257 field \citep{tes95}, and there do not appear to be 
many cool (metal-rich) red giants that would have been missed in the 
spectroscopic survey.  Olszewski (1993) state that many of their blue targets
turned out to be Galactic field stars.  Selection bias is therefore unlikely
to account for the 
large difference in MDF between the inner and outer field.

Our preferred explanation for the difference is that we are seeing
an effect analogous to that described by Bica et al. (1999) for the star
clusters.  The intermediate-age, metal-rich stars are more centrally
concentrated than the older, metal-poor stars.  Hence for a given
sample size, the relative fraction of metal-poor stars rises with
increasing radius.  However, the intermediate-age stars throughout
the LMC disk have apparently been enriched to the same level.  It is 
evident that spectroscopic studies of the inner LMC disk must contain 
$\gtrsim$100 stars if the metal-poor component seen by Olszewski (1993)
in the outer disk is to be reliably detected.
 
The similarity of field stars to star clusters is reinforced
by comparison to the cluster abundance study of Olszewski et al (1991).
Their abundance histogram has been transformed to the Carretta \& Gratton (1997)
metallicity scale by the same procedure we applied to the
field star data.  The two samples are plotted in Figure \ref{o9193fig};
It is seen that our Disk 1 field, which is dominated by stars with 
[Fe/H] $\approx$ $-$0.55, resembles the abundance distribution of
intermediate-age star clusters.  However, a metal-poor, globular
cluster-like population is found to be lacking in the inner disk.

Attention is also drawn to the excess numbers of field stars with
abundances between $-$1.2 $\leq$ [Fe/H] $\leq$ $-$0.8, relative to
the star cluster population.  This is direct evidence that the star-formation
in the field was not as quiescent as cluster formation, during the 4--12
Gyr age gap, as inferred from WFPC2 color-magnitude diagrams
\citep{gal99,hol99}.  Although the absolute numbers of stars are small,
and hence statistically uncertain, Figure \ref{o9193fig} may be taken to 
support the idea that field star formation
did the bulk of the chemical enrichment work in the LMC during the cluster
dark ages, as required by the cluster abundance distribution \citep{ols96}.
Larger spectroscopic samples are urgently needed in order to put this 
result on firmer statistical ground.  However, it is already supported
by our corrected photometric MDF (Figure \ref{cpmdf}), which shows an
excess of stars at [M/H]$'$ $\approx$ $-$0.95 compared to the best-fit
Gaussian that matches the majority stellar population.

\section{Summary and Future Work \label{sumsec}}

In this paper, we have described the first results of a program to
measure the metallicity distribution functions (MDF) of red giant
stars in selected LMC fields.  The specific fields coincide
with regions of deep WFPC2 photometry, so that color-magnitude diagram
and abundance information may be coupled to derive an accurate and
unique history of star-formation for the bar and disk of the LMC. 

The data reported here were taken in a single field of the inner 
LMC disk, 1$\fdg$8 southwest of the optical center of the bar.
Two techniques for the derivation of abundances are employed:
medium resolution spectroscopy at the \caii\ IR triplet, and
intermediate-band photometry in the Str\"{o}mgren system.
Str\"{o}mgren
photometry for the remainder of our WFPC2 fields is in hand, and we 
plan to obtain much larger spectroscopic samples in forthcoming 
observing seasons.  

The possible effects of systematic errors on the \caii\ triplet 
abundance scale as applied to the LMC were examined.  Contamination
of the sample by AGB stars was shown to have a negligible effect on
our derived MDF.   The physics of \caii\ line formation suggest that
the metal-rich end of our MDF ([Fe/H] $\gtrsim$ $-$0.4 dex) might be
biased to low values by $\approx$0.1 dex; however, the observational
method to correct for surface gravity effects, based on the magnitude
of the horizontal branch (red clump) probably compensates for this
to some degree.  A potentially large, but unconstrained, source of
error is the variation in [Ca/Fe] ratio among LMC stars.  However,
none of these systematics is expected to greatly affect the metal-poor
end of our MDF, which is where the most striking comparison to 
other work can be made, and where the strongest constraints can
be placed on \sfh\ models.

Systematic effects in the \caii\ triplet may change the 
details of the high-metallicity end of our MDF, but its basic form
is robust.  However, we note that the \caii\ triplet method as it
stands should only be applied to intermediate-age and young systems,
where [Fe/H] may exceed $-$0.3, with extreme caution.  We have 
embarked on a new study to addess this issue and derive a new
calibration of the \caii\ triplet from a sample of open and globular 
clusters.

We consider the effects of interstellar reddening on the 
Str\"{o}mgren colors; a wide range of behaviors reported
in the literature, and we rederive the reddening relations.
Although the mean reddening in our Disk 1 field
is small, caution must be exercised in applying the results of 
Str\"{o}mgren photometry without careful attention to reddening.
Despite this weakness, the insensitivity of Str\"{o}mgren
colors to age makes it the preferred photometric abundance indicator
where large age variations are known to be present.

The MDF determined photometrically is in good agreement with
our spectroscopic results, although there is a significant 
offset between the two techniques that is linearly dependent
on [M/H].  The scatter in the correlation relation is not
larger than predicted based on the random errors in each 
technique.  The origin of this offset should be elucidated by
high-resolution spectroscopy, in combination with further
photometry of stars on the lower RGB, and for metal-poor stars.

We find that the 
statistical properties of the spectroscopic MDF are well-preserved 
in the photometric sample, with the following caveats:  {\it 1)} 
foreground stars are an important contributor to the metal-rich wing
of the photometric MDF; {\it 2)} the scatter of individual measurements
about the mean color-metallicity relation must be taken into account
when assessing the width of the abundance distribution, and {\it 3)}
the lack of low-metallicity stars in our spectroscopic sample makes
the calibration of Str\"{o}mgren abundances to the \caii\ triplet
scale quite uncertain below [Fe/H] $\approx$ $-$1.1.  However,
insofar as the offset between abundance scales is stable for 
intermediate-age stars, Str\"{o}mgren photometry can be an important
tool for exploring the MDF of large numbers of stars.

Based primarily on our spectroscopic data,
we find the median metallicity of the inner LMC disk to be [Fe/H] = 
$-$0.57 $\pm$0.04, while the mean [Fe/H] = $-$0.64 $\pm$0.02.  The 
errors are random only, and do not reflect the possibility of systematic
errors at the $\pm$0.1--0.2 dex level.  There is a tail of metal-poor
stars in the spectroscopic MDF that extends down to at least [Fe/H] =
$-$1.6; the indentification of this tail with a possible metal-poor 
component of the photometric MDF, containing $\lesssim$10\% of the stars, is 
not certain.  This component of the photometric MDF, when corrected
via Equation \ref{coreqn} to the \caii\ triplet scale, shows a mean
[M/H]$'$ = $-$0.93.
Larger spectroscopic samples are required to address
this issue.  The intrinsic dispersion in abundances is high, $\approx$0.25
dex; this likely reflects a range in ages as well as a scatter about
the mean age-metallicity relation.
Half of the red giants in our Disk 1 field fall in the 
range $-$0.83 $\leq$ [Fe/H] $\leq$ $-$0.41.

We find strong evidence for a radial variation in the relative
fraction of metal-rich and metal-poor stars, by comparison to the
work of Olszewski (1993).  Their outer field shows 3 times as many 
metal-poor ([Fe/H] $\leq$ $-$1) stars as the Disk 1 field, although the location of
the high-metallicity peak at [Fe/H] $\approx$ $-$0.6 is similar.
This is interpreted as supporting a model
in which stars of a given age share similar abundances, but the
intermediate-age stars are more centrally concentrated than the
older stars.  This would extend to the field populations a trend
previously noted for the star clusters \citep{bic99}.  

By relative number, the Disk 1 field shows twice as many stars
in the range $-$1.3 $\leq$ [Fe/H] $\leq$ $-$0.8 than the star
cluster sample studied by Olszewski et al. (1991).  This may be direct evidence
for the hypothesis that star formation in the field did most of the
chemical evolution work in the LMC during the time period
from 4--12 Gyr ago \citep{ols96,gal99}.  Support for this comes
from the photometric MDF, which, while imprecise, shows little evidence
for a gap at [M/H] $\sim$ $-$1.  More work is needed 
to confirm this result, due to the small number of stars observed 
spectroscopically.


Future work will see the application of these results to 
the detemination of the \sfh\ of the LMC disk.
We have obtained deep WFPC2 photometry for two LMC fields, one 
in the inner disk and one at the center of the bar \citep{sme99}.
Further data will be obtained for a field at the western end of the
bar in STScI Cycle 9.
These data will result in the best color-magnitude diagrams ever
constructed for an extragalactic system.  From these CMDs, we will
attempt to derive the \sfh\ of the LMC bar and disk by quantitative
comparison to synthetic CMDs (e.g., Tolstoy \& Saha 1996).  
The data we have obtained here will be used to place additional 
conditions on the synthetic CMDs: not only must the ``best fit''
model match the observed distribution of points in color and magnitude,
we will require that the synthetic MDF also matches the observed one.
A full analysis of our WFPC2 CMDs, which contain an order of magnitude
more stars than similar data in the literature, will be reported in a forthcoming paper 
\citep{sme00}.  

\acknowledgments

This paper is adapted from Chapter 5 of the Ph.D. dissertation of
Andrew A. Cole in the Astronomy Department of the University of 
Wisconsin-Madison.  We are very pleased to thank the support staff
at CTIO for their invaluable assistance in conducting these observations.
AAC would like to thank his thesis advisors,
Jay Gallagher and Tammy Smecker-Hane.  
We would like to thank the anonymous referee for calling our
attention to an error in Figure 14 which clouded our 
interpretation of the correlation between spectroscopic and
photometric abundance determinations.  Travel support for AAC was
provided by the Cerro Tololo Interamerican Observatory.
Financial support was provided in part by the NSF through grant
AST-9619460 to TSH, and 
by NASA through grant GO\#7382
from the Space Telescope Science Institute, which is operated by the
Association of Universities for Research in Astronomy, Inc. under NASA
contract NAS-26555.  
The WFPC2 Investigation Definition Team provided partial support 
during the completion of AAC's thesis.   Additional support provided
to JSG by a grant from the Vilas Trust to the University of Wisconsin-Madison.
This research has made use of: NASA's
Astrophysics Data System Abstract Service; 
the Digitized Sky Survey, which was produced at the Space Telescope
Science Institute under U.S. Government grant NAG W-2166; and
the Canadian Astronomy Data Centre, which is operated by the 
Herzberg Institute of Astrophysics, National Research Council of
Canada.  AAC would like to thank Arthur Guinness Son \& Co., Ltd.
for many pints
of life-sustaining velvety black nectar.

\clearpage

\renewcommand{\arraystretch}{.6}

\begin{landscape}

\begin{deluxetable}{ccccccccccc}
\tablewidth{0pt}
\tablecaption{Identification of Program Stars \label{datatab}}
\tablehead{
\colhead{ID} &
\colhead{Number} &
\colhead{$\alpha$} &
\colhead{$\delta$} &
\colhead{$\Sigma$W(Ca)} &
\colhead{$\sigma$W} &
\colhead{V} &
\colhead{$\sigma$V} &
\colhead{[Fe/H]} &
\colhead{$\sigma$[Fe/H]} &
\colhead{Subgroup} \\ 
\colhead{} &
\colhead{} &
\colhead{(J2000.0)} &
\colhead{(J2000.0)} &
\colhead{(\AA)} &
\colhead{(\AA)} &
\colhead{(mag)} &
\colhead{(mag)} &
\colhead{(dex)} &
\colhead{(dex)} &
\colhead{}}
\startdata
1 & 0150-02697287 & 05:11:35.934 & -71:13:32.49 & 5.978 & 0.289 & 17.133 & 0.012 & -0.792 & 0.140 & {\it ii} \\
2 & 0150-02699460 & 05:11:40.558 & -71:13:32.63 & 5.541 & 0.233 & 17.632 & 0.007 & -0.843 & 0.122 & {\it ii} \\
3 & 0150-02677898 & 05:10:56.610 & -71:16:42.30 & 5.742 & 0.230 & 17.808 & 0.007 & -0.719 & 0.121 & {\it ii} \\
4 & 0150-02701335 & 05:11:44.596 & -71:07:18.21 & 7.317 & 0.284 & 17.144 & 0.006 & -0.251 & 0.138 & {\it ii} \\
5 & 0150-02701911 & 05:11:45.833 & -71:07:31.69 & 7.185 & 0.251 & 16.707 & 0.006 & -0.413 & 0.128 & {\it iii} \\
8 & 0150-02692129 & 05:11:24.810 & -71:16:25.77 & 6.450 & 0.261 & 17.909 & 0.011 & -0.409 & 0.131 & {\it ii} \\
9 & 0150-02691174 & 05:11:22.924 & -71:16:59.02 & 7.106 & 0.280 & 17.410 & 0.008 & -0.270 & 0.137 & {\it ii} \\ 
10 & 0150-02690283 & 05:11:21.244 & -71:17:20.87 & 5.759 & 0.247 & 17.205 & 0.017 & -0.862 & 0.127 & {\it iv} \\
11 & 0150-02689820 & 05:11:20.342 & -71:17:34.68 & 6.906 & 0.283 & 17.437 & 0.009 & -0.343 & 0.138 & {\it ii} \\
13 & 0150-02686676 & 05:11:14.372 & -71:18:58.24 & 6.180 & 0.265 & 17.260 & 0.011 & -0.679 & 0.132 & {\it ii} \\
15 & 0150-02684855 & 05:11:10.930 & -71:14:30.44 & 6.662 & 0.205 & 17.398 & 0.010 & -0.451 & 0.114 & {\it ii} \\
16 & 0150-02682408 & 05:11:06.094 & -71:14:48.70 & 6.292 & 0.253 & 17.702 & 0.010 & -0.524 & 0.128 & {\it ii} \\
17 & 0150-02679627 & 05:11:00.480 & -71:15:09.98 & 5.211 & 0.243 & 18.383 & 0.039 & -0.789 & 0.126 & {\it i} \\
18 & 0150-02676927 & 05:10:54.463 & -71:15:33.72 & 5.497 & 0.234 & 18.091 & 0.006 & -0.747 & 0.122 & {\it i} \\
19 & 0150-02655815 & 05:10:06.601 & -71:15:50.91 & 6.432 & 0.257 & 17.188 & 0.005 & -0.596 & 0.130 & {\it ii} \\
20 & 0150-02655172 & 05:10:05.348 & -71:15:29.88 & 6.592 & 0.287 & 17.230 & 0.005 & -0.521 & 0.139 & {\it ii} \\
21 & 0150-02654809 & 05:10:04.592 & -71:15:13.82 & 6.406 & 0.245 & 17.356 & 0.006 & -0.564 & 0.126 & {\it ii} \\
23 & 0150-02716108 & 05:12:16.561 & -71:08:56.01 & 3.957 & 0.249 & 17.358 & 0.022 & -1.548 & 0.127 & {\it ii} \\
24 & 0150-02715410 & 05:12:15.075 & -71:08:35.74 & 4.461 & 0.206 & 17.561 & 0.022 & -1.295 & 0.114 & {\it iv} \\
25 & \nodata \tablenotemark{a} & 05:12:10 & -71:07:33 & 4.357 & 0.173 & 18.491 & 0.023 & -1.105 & 0.105 & {\it i} \\
26 & 0150-02711752 & 05:12:07.264 & -71:06:56.33 & 6.582 & 0.254 & 18.244 & 0.022 & -0.272 & 0.129 & {\it i} \\
29 & 0150-02680078 & 05:11:01.504 & -71:14:37.22 & 4.617 & 0.378 & 18.548 & 0.006 & -0.987 & 0.171 & {\it i,iv} \\
30 & \nodata \tablenotemark{a} & 05:11:03 & -71:14:18 & 6.009 & 0.244 & 17.517 & 0.012 & -0.684 & 0.126 & {\it ii} \\
31 & 0150-02681765 & 05:11:04.802 & -71:13:49.78 & 5.753 & 0.253 & 18.377 & 0.007 & -0.573 & 0.128 & {\it i,iv} \\
33 & 0150-02707271 & 05:11:57.508 & -71:09:09.61 & 6.096 & 0.249 & 17.830 & 0.018 & -0.571 & 0.127 & {\it ii} \\
35 & 0150-02702958 & 05:11:48.130 & -71:12:16.18 & 5.232 & 0.240 & 17.858 & 0.019 & -0.911 & 0.124 & {\it ii} \\
36 & 0150-02698027 & 05:11:37.565 & -71:12:29.08 & 3.135 & 0.261 & 18.778 & 0.006 & \nodata & \nodata & TiO \\
37 & 0150-02696795 & 05:11:34.945 & -71:12:30.62 & 5.371 & 0.264 & 17.534 & 0.007 & -0.936 & 0.132 & {\it ii} \\
38 & 0150-02694661 & 05:11:30.385 & -71:12:35.73 & 6.707 & 0.291 & 16.969 & 0.006 & -0.540 & 0.141 & {\it iii} \\
39 & 0150-02658835 & 05:10:12.748 & -71:10:28.32 & 5.820 & 0.254 & 17.497 & 0.005 & -0.765 & 0.129 & {\it ii} \\
40 & 0150-02658096 & 05:10:11.254 & -71:10:16.82 & 6.812 & 0.258 & 17.651 & 0.005 & -0.328 & 0.130 & {\it ii} \\
41 & 0150-02656626 & 05:10:08.289 & -71:09:58.03 & 5.229 & 0.299 & 17.191 & 0.005 & -1.079 & 0.143 & {\it ii} \\
42 & 0150-02665754 & 05:10:27.883 & -71:18:59.64 & 7.343 & 0.199 & 17.074 & 0.006 & -0.258 & 0.112 & {\it iii} \\
44 & 0150-02646913 & 05:09:47.376 & -71:14:11.09 & 7.103 & 0.277 & 16.840 & 0.005 & -0.413 & 0.136 & {\it iii} \\
45 & 0150-02648095 & 05:09:50.039 & -71:13:36.77 & 6.763 & 0.233 & 17.500 & 0.005 & -0.385 & 0.122 & {\it ii} \\
46 & 0150-02717484 & 05:12:19.480 & -71:19:56.67 & 6.837 & 0.226 & 16.903 & 0.022 & -0.504 & 0.120 & {\it iv} \\
47 & 0150-02715689 & 05:12:15.655 & -71:19:55.29 & 6.704 & 0.266 & 17.248 & 0.021 & -0.472 & 0.133 & {\it ii} \\
49 & 0150-02689598 & 05:11:19.900 & -71:16:39.58 & 7.017 & 0.254 & 16.936 & 0.013 & -0.423 & 0.129 & {\it iii} \\
50 & 0150-02687605 & 05:11:16.096 & -71:16:40.23 & 6.889 & 0.222 & 17.324 & 0.009 & -0.378 & 0.119 & {\it iii} \\
51 & 0150-02685571 & 05:11:12.266 & -71:16:39.77 & 5.921 & 0.279 & 17.080 & 0.009 & -0.828 & 0.137 & {\it ii} \\
\enddata
\tablenotemark{a}{No match found in USNO-A catalog within 2$\arcsec$.}
\tablecomments{ID numbers and coordinates are taken from the USNO-A astrometric
reference catalog \citep{mon98}, except where noted.  Magnitudes are from Cole (1999).
For subgroup definitions, see text. Star 36 is excluded from the abundance analysis
because of contamination by TiO bands.}
\end{deluxetable}

\end{landscape}

\clearpage

\begin{deluxetable}{lcccr}
\tablewidth{0pt}
\tablecaption{Calibrating Clusters \label{globref}}
\tablehead{
\colhead{Cluster} &
\colhead{V$_{\mathrm{HB}}$} &
\colhead{[Fe/H]$_{\mathrm{ZW84}}$} &
\colhead{[Fe/H]$_{\mathrm{CG97}}$} &
\colhead{[Ca/Fe]}}
\startdata
M 68 (NGC 4590)  & 15.68 & $-$2.09 $\pm$0.11 & $-$2.00 $\pm$0.03 & $+$0.32 \\
NGC 1904        & 16.15 & $-$1.68 $\pm$0.09 & $-$1.37 $\pm$0.05 & $+$0.30 \\
NGC 288         & 15.30 & $-$1.40 $\pm$0.12 & $-$1.14 $\pm$0.03 & $+$0.12 \\
NGC 1851        & 16.09 & $-$1.33 $\pm$0.09 & $-$1.03 $\pm$0.06 & $+$0.03 \\
47 Tuc (NGC 104) & 14.06 & $-$0.71 $\pm$0.08 & $-$0.78 $\pm$0.02 & $-$0.02 \\
M 67 (NGC 2682)  & 10.56\tablenotemark{a} &
\multicolumn{2}{c} {$-$0.06 $\pm$0.07\tablenotemark{b}} & 0.00 \\
\enddata
\tablenotetext{a}{from Sarajedini 1998.}
\tablenotetext{b}{from Nissen et al. 1987.}
\tablecomments{Horizontal branch magnitudes from the online catalog
of Harris (1996), except as noted.  [Fe/H] values from Zinn \& West (1984) and
Carretta \& Gratton (1997), except as noted. [Ca/Fe] values from Carney (1996).}
\end{deluxetable}

\clearpage

\begin{deluxetable}{lccc}
\tablewidth{0pt}
\tablecaption{Calibrator Magnitudes and Equivalent Widths \label{globcal}}
\tablehead{
\colhead{Star} &
\colhead{V$-$V$_{\mathrm{HB}}$ (mag)} &
\colhead{$\Sigma$W(Ca)\tablenotemark{a} \, (\AA)} &
\colhead{Notes}}
\startdata
M 68 / I-74	& $-$1.01 & 2.178 $\pm$0.108 & 1,2 \\
M 68 / I-256	& $-$2.96 & 3.265 $\pm$0.149 & 1,2 \\
M 68 / II-28	& $-$1.88 & 2.715 $\pm$0.134 & 1,2 \\
\tableline
NGC 1904 / 006	& $-$0.89 & 4.113 $\pm$0.164 & 2,3 \\
NGC 1904 / 015	& $-$2.97 & 5.281 $\pm$0.164 & 3,4 \\
NGC 1904 / 089	& $-$1.49 & 4.408 $\pm$0.133 & 2,3 \\
NGC 1904 / 240	& $-$2.20 & 4.833 $\pm$0.171 & 3,4 \\
NGC 1904 / 241	& $-$2.58 & 4.818 $\pm$0.144 & 3,4 \\
\tableline
NGC 288 / C20	& $-$2.30 & 5.109 $\pm$0.173 & 2,5 \\
NGC 288 / C23	& $-$0.32 & 3.954 $\pm$0.149 & 2,5 \\
NGC 288 / C36	& $-$1.43 & 4.357 $\pm$0.194 & 2,5 \\
NGC 288 / C41	& $-$0.82 & 4.299 $\pm$0.190 & 2,5 \\
\tableline
NGC 1851 / 109	& $-$1.17 & 4.725 $\pm$0.187 & 6,7 \\
NGC 1851 / 112	& $-$2.22 & 5.640 $\pm$0.165 & 6,8 \\
NGC 1851 / 293	& $-$0.59 & 4.705 $\pm$0.296 & 6,7 \\
NGC 1851 / 294	& $-$2.61 & 5.886 $\pm$0.137 & 6,9 \\
NGC 1851 / 333	& $-$2.01 & 5.453 $\pm$0.114 & 6,8 \\
\tableline
47 Tuc / L1523	& $-$0.09 & 4.488 $\pm$0.233 & 10,11 \\
47 Tuc / L2506	& $-$0.00 & 4.787 $\pm$0.214 & 10,11 \\
47 Tuc / L2514	& $+$0.02 & 4.442 $\pm$0.147 & 10,11 \\
47 Tuc / L5406	& $-$1.23 & 5.275 $\pm$0.227 & 9,10 \\
47 Tuc / L5422	& $-$1.59 & 5.545 $\pm$0.156 & 9,10 \\
\tableline
M 67 / F84	& $+$0.04 & 5.790 $\pm$0.199 & 12,13 \\
M 67 / F105	& $-$0.23 & 5.700 $\pm$0.162 & 12,13 \\
M 67 / F108	& $-$0.83 & 6.092 $\pm$0.166 & 12,13 \\
M 67 / F141	& $-$0.09 & 5.843 $\pm$0.149 & 12,13 \\
M 67 / F170	& $-$0.83 & 5.943 $\pm$0.223 & 12,13 \\
\enddata
\tablenotetext{a}{Summed equivalent widths of $\lambda \lambda$
8542, 8662 lines.}
\tablerefs{(1) Identification from Harris 1975;
(2) Photometry from Suntzeff et al. 1993;
(3) ID: Stetson \& Harris 1977;
(4) Photometry: Stetson \& Harris 1977;
(5) ID: Cannon 1974;
(6) ID: Stetson 1981;
(7) Photometry: Stetson 1981;
(8) Photometry: Armandroff \& Da Costa 1991;
(9) Photometry: Da Costa \& Armandroff 1995;
(10) ID: Lee 1977;
(11) Photometry: Lee 1977;
(12) ID: Johnson \& Sandage 1955;
(13) Photometry: Janes \& Smith 1984.}
\end{deluxetable}

\clearpage

\begin{deluxetable}{lcccll}
\tablewidth{0pt}
\tablecaption{Selected LMC Fields \label{fields}}
\tablehead{
\colhead{Field} &
\colhead{$\alpha$ (J2000.0)} &
\colhead{$\delta$ (J2000.0)} &
\colhead{$\rho$\tablenotemark{a} \, (deg.)} &
\colhead{Notes} &
\colhead{References}}
\startdata
Disk 1 A$+$B & 5:10:50 & $-$71:13:31 & 1.8 & Inner disk, near SL336 & 1,2 \\
HST-1 & 5:14:44 & $-$65:17:43 & 4.5 & Outer disk, near NGC 1866 & 3,4,5,6,7 \\
HST-2 & 5:23:51 & $-$69:49:28 & 0.1 & Bar, near HS-275 & 1,2,7 \\
HST-3 & 5:58:21 & $-$68:21:19 & 3.4 & Outer disk, near Hodge 10 & 6,7 \\
\enddata
\tablenotetext{a}{Distance from the optical center of the bar
\citep{dev72}, on the sky.}
\tablerefs{WFPC2 color-magnitude diagram studies:
(1) Smecker-Hane et al. 1999; (2) Cole et al. 1999; (3) Gallagher et al.
1996; (4) Holtzman et al. 1997; (5) Stappers et al. 1997; (6) Geha et al.
1998; (7) Holtzman et al. 1999.}
\end{deluxetable}

\clearpage

\begin{deluxetable}{lccl}
\tablewidth{0pt}
\tablecaption{Catalog of Photometric Observations \label{obslog1}}
\tablehead{
\colhead{Field} &
\colhead{Date} &
\colhead{Seeing (arcsec)} &
\colhead{Exposures}}
\startdata
Disk 1A	& 1997 Nov 28	& 1$\farcs$4 & 1200s $y$ \\
Disk 1A	& 1997 Nov 28	& 1$\farcs$9 & 1300s $b$ \\
Disk 1A	& 1997 Nov 28	& 1$\farcs$5 & 2$\times$ 1300s $v$ \\
Disk 1A	& 1997 Nov 28	& 1$\farcs$4 & 1300s $v$ \\
Disk 1A	& 1997 Nov 28	& 1$\farcs$6 & 1300s $b$ \\
Disk 1A	& 1997 Nov 28	& 1$\farcs$3 & 1200s $y$ \\
\tableline
Disk 1A	& 1997 Dec 1	& 1$\farcs$4 & 1300s $v$ \\
\enddata
\end{deluxetable}

\clearpage

\begin{deluxetable}{rrrrrrrrr}
\tablewidth{0pt}
\tablecaption{Str\"{o}mgren Photometry of the Disk 1A Field
\tablenotemark{a} \label{photab}}
\tablehead{
\colhead{ID} &
\colhead{$\Delta \delta$ ($\arcsec$)\tablenotemark{b}} &
\colhead{$\Delta \alpha$ ($\arcsec$)\tablenotemark{c}} &
\colhead{$y$} &
\colhead{$\sigma _y$} &
\colhead{$b-y$} &
\colhead{$\sigma _{by}$} &
\colhead{$m_1$} &
\colhead{$\sigma _{m1}$}}
\startdata
200001 & $-$0.274 & 388.203 & 20.184 & 1.444 & 0.182 & 1.865 & $-$0.021 & 2.360 \\
201648 & $-$0.083 & 165.561 & 19.059 & 0.287 & 0.889 & 0.362 &  0.206 & 0.457 \\
201495 &  0.000 & 149.296 & 19.550 & 0.154 & 0.219 & 0.196 & $-$0.020 & 0.249 \\
202309 &  0.078 & 264.597 & 19.607 & 0.043 & 0.176 & 0.063 &  0.157 & 0.092 \\
 32523 &  0.087 & 147.713 & 19.256 & 0.031 & 0.498 & 0.050 &  0.300 & 0.079 \\
\enddata
\tablenotetext{a}{The complete version of this table is in the electronic
verstion of the Journal.  The printed edition contains only a sample.}
\tablenotetext{b}{Offset north of $\delta$ = $-$71:20:56 (J2000.0).}
\tablenotetext{c}{Offset west of $\alpha$ = 5:33:54 (J2000.0).}
\end{deluxetable}


\begin{thebibliography}{}

\bibitem[Allen 1973]{all73} Allen, C.W. 1973, ``Astrophysical Quantities'',
	3rd. ed. (London: Athlone Press)
\bibitem[Alves \& Sarajedini 1999]{alv99} Alves, D.R. \& Sarajedini,
	A. 1999, \apj, 511, 225
\bibitem[Anthony-Twarog et al. 1995]{ant95} Anthony-Twarog, B.J., 
	Twarog, B.A., \& Craig, J. 1995, \pasp, 107, 32
\bibitem[Aparicio et al. 1996]{apa96} Aparicio, A., Gallart, C., Chiosi, C.,
	\& Bertelli, G. 1996, \apjl, 469, L97
\bibitem[Armandroff \& Zinn 1988]{arm88} Armandroff, T.E., \& Zinn, R.
	1988, \aj, 96, 92
\bibitem[Armandroff \& Da Costa 1991]{arm91} Armandroff, T.E., \&
	Da Costa, G.S. 1991, \aj, 101, 1329
\bibitem[Baade 1963]{baa63} Baade, W. 1963, ``Evolution of Stars and 
	Galaxies'', ed. C. Payne-Gaposchkin, (Harvard Univ. Pr.: Cambridge)
\bibitem[Bahcall \& Soneira 1980]{bah80} Bahcall, J.N., \& Soneira,
	R.M. 1980, \apjs, 44, 73
\bibitem[Bell \& Gustafsson 1978]{bel78} Bell, R.A., \& Gustafsson,
	B. 1978, \aaps, 34, 229
\bibitem[Bertelli et al. 1992]{ber92} Bertelli, G., Mateo, M., Chiosi, C.,
	\& Bressan, A. 1992, \apj, 488, 400
\bibitem[Bica \& Alloin 1987]{bic87} Bica, E., \& Alloin, D.M. 1987,
	\aap, 186, 49
\bibitem[Bica et al. 1999]{bic99} Bica, E., Schmitt, H.R., Dutra, C.M.,
	\& Oliveira, H.L. 1999, \aj, 117, 238
\bibitem[Bond 1980]{bon80} Bond, H.E. 1980, \apjs, 44, 517
\bibitem[Boothroyd \& Sackman 1999]{boo99} Boothroyd, A.I.,
	\& Sackmann, I.-J. 1999, \apj, 474, 188
\bibitem[Butcher 1977]{but77} Butcher, H. 1977, \apj, 216, 372
\bibitem[Cannon 1974]{can74} Cannon, R.D. 1974, \mnras, 167, 551 
\bibitem[Cardelli et al. 1989]{car89} Cardelli, J.A., Clayton, G.C.,
	\& Mathis, J.S. 1989, \apj, 345, 245
\bibitem[Carney 1996]{car96} Carney, B.W. 1996, \pasp, 108, 900
\bibitem[Carretta \& Gratton 1997]{car97} Carretta, E., \& Gratton,
	R.G. 1997, \aaps, 121, 95
\bibitem[Charbonnel 1994]{cha94} Charbonnel, C. 1994, \aap, 282, 811
\bibitem[Cole 1999]{col99b} Cole, A.A. 1999, Ph.D. Thesis, University of
	Wisconsin--Madison.
\bibitem[Cole et al. 1999]{col99c} Cole, A.A., Smecker-Hane, T.A.,
	\& Gallagher, J.S., III 1999, \baas, \#08.15
\bibitem[Crawford \& Barnes 1970]{cra70} Crawford, D.L., \& Barnes,
	J.V. 1970, \aj, 75, 978
\bibitem[Crawford \& Mandwewala 1976]{cra76} Crawford, D.L., \&
	Mandwewala, N. 1976, \pasp, 88, 917
\bibitem[Da Costa 1991]{dac91} Da Costa, G.S. 1991, in IAU Symp. 148,
	``The Magellanic Clouds'', eds. R. Haynes \& D. Milne,
	(Kluwer: Dordrecht), p. 183
\bibitem[Da Costa 1998]{dac98b} Da Costa, G.S. 1998, in 
	``Stellar Astrophysics for the Local Group'', eds.
	A. Aparicio, A. Herrero, \& F. S\'{a}nchez, (Cambridge:
	Cambridge Univ. Pr.), p. 351
\bibitem[Da Costa 1999]{dac99} Da Costa, G.S. 1999, in IAU Symp. 190,
	``New Views of the Magellanic Clouds'', eds. Y.-H. Chu, N.B.
        Suntzeff, J.E. Hesser, \& D.A. Bohlender, (San Francisco: PASP),
        p. 397
\bibitem[Da Costa \& Armandroff 1990]{dac90} Da Costa, G.S.,
	\& Armandroff, T.E. 1990, \aj, 100, 162
\bibitem[Da Costa \& Armandroff 1995]{dac95} Da Costa, G.S., 
	\& Armandroff, T.E. 1995, \aj, 109, 2533
\bibitem[Da Costa \& Hatzidimitriou 1998]{dac98} Da Costa, G.S.,
	\& Hatzidimitriou, D. 1998, \aj, 115, 1934
\bibitem[de Freitas Pacheco et al. 1998]{def98} de Freitas Pacheco,
	J.A., Barbuy, B., \& Idiart, T. 1998, \aap, 332, 19
\bibitem[Delgado et al. 1997]{del97} Delgado, A.J., Alfaro, E.J., \&
	Cabrera-Ca\~{n}o, J. 1997, \aj, 113, 713
\bibitem[de Vaucoleurs \& Freeman 1972]{dev72} de Vaucouleurs, G., \&
	Freeman, K.C. 1972, Vistas Astron., 14, 163
\bibitem[D\'{\i}az et al. 1989]{dia89} D\'{\i}az, A.I., Terlevich,
	E., \& Terlevich, R. 1989, \mnras, 239, 325
\bibitem[Dopita et al. 1997]{dop97} Dopita, M.A., Vassiliadis, E., 
	et al. 1997, \apj, 474, 188
\bibitem[Edvardsson et al. 1993]{edv93} Edvardsson, B., Andersen, J.,
	Gustaffson, B., Lambert, D.L., Nissen, P.E., \& Tomkin, J.
	1993, \aap, 275, 101
\bibitem[Faber \& French 1980]{fab80} Faber, S.M., \& French, H.B. 1980,
	\apj, 235, 405
\bibitem[Fitzpatrick 1999]{fit99} Fitzpatrick, E.L. 1999, \pasp, 111, 63
\bibitem[Frogel et al. 1990]{fro90} Frogel, J.A., Mould, J.R.,
	\& Blanco, V.M. 1990, \apj, 352, 96
\bibitem[Gallagher et al. 1996]{gal96} Gallagher, J.S., III, Mould, J.R.,
	de Feijter, E., Holtzman, J., Stappers, B., Watson, A., 
	Trauger, J., \& The WFPC2 IDT 1996, \apj, 466, 732
\bibitem[Gallagher et al. 1998]{gal98} Gallagher, J.S., III, Tolstoy, E.,
	Dohm-Palmer, R.C., Skillman, E.D., Cole, A.A., Hoessel, J.G.,
	Saha, A., \& Mateo, M. 1998, \aj, 115, 1869
\bibitem[Gallagher et al. 1999]{gal99} Gallagher, J.S., III, Cole, A.A.,
	Holtzman, J.A., \& Smecker-Hane, T.A. 1999, in IAU Symp. 190,
	``New Views of the Magellanic Clouds'', eds. Y.-H. Chu, N.B.
	Suntzeff, J.E. Hesser, \& D.A. Bohlender, (San Francisco: PASP),
	p. 306 
\bibitem[Gascoigne 1964]{gas64} Gascoigne, S.C.B. 1964, in IAU-USRI
	Symp. 20, ``The Galaxy and the Magellanic
        Clouds'', eds. F.J. Kerr \& A.W. Rodgers,
        (Austr. Acad. Sci: Canberra), p. 354
\bibitem[Geha et al. 1998]{geh98} Geha, M.C., et al. 1998, \aj, 115, 1045
\bibitem[Gilmore \& Wyse 1991]{gil91} Gilmore, G., \& Wyse, R.F.G. 1991,
	\apj, 367, L55
\bibitem[Girardi et al. 2000]{gir00} Girardi, L., Bressan, A., 
	Bertelli, G., \& Chiosi, C. 2000, \aaps, 141, 371
\bibitem[Grebel \& Richtler 1992]{gre92} Grebel, E.K., \& Richtler,
	T. 1992, \aap, 253, 359
\bibitem[Hardy et al. 1984]{har84} Hardy, E., Buonanno, R., Corsi, C.E.,
	Janes, K.A., \& Schommer, R.A. 1984, \apj, 278, 592
\bibitem[Harris 1975]{har75} Harris, W.E. 1975, \apjs, 29, 397
\bibitem[Harris 1996]{har96} Harris, W.E. 1996, \aj, 112, 1487
\bibitem[Hilker 2000]{hil99} Hilker, M. 1999, \aap, in press,
	astro-ph/9911387
\bibitem[Hill et al. 1995]{hil95} Hill, V., Andrievsky, S., \& 
	Spite, M. 1995, \aap, 293, 347
\bibitem[Hodge 1989]{hod89} Hodge, P.W. 1989, \araa, 27, 139
\bibitem[Holtzman et al. 1997]{hol97} Holtzman, J.A., Mould, J.R.,
	Gallagher, J.S., III, Watson, A.M., Grillmair, C.J., \& 
	The WFPC2 IDT 1997, \aj, 113, 656
\bibitem[Holtzman et al. 1999]{hol99} Holtzman, J.A., Gallagher, J.S.,
	III, Cole, A.A., Mould, J.R., Grillmair, C.J., \& The WFPC2
	IDT 1999, \aj, 118, 2262
\bibitem[Holtzman et al. 1999b]{hol99b} Holtzman, J.A., Mould, J.R.,
	\& Gallagher, J.S., III 1999b, in IAU Symp. 190, 
	``New Views of the Magellanic Clouds'', eds. Y.-H. Chu, N.B.
        Suntzeff, J.E. Hesser, \& D.A. Bohlender, (San Francisco: PASP),
        p. 306
\bibitem[Iben 1968]{ibe68} Iben, I., Jr. 1968, \nat, 220, 143
\bibitem[Isserstedt 1975]{iss75} Isserstedt, J. 1975, \aaps, 19, 259
\bibitem[Janes 1984]{jan84b} Janes, K.A. 1984, \pasp, 96, 977
\bibitem[Janes \& Smith 1984]{jan84a} Janes, K.A., \& Smith, G.H. 1984,
	\aj, 89, 487 
\bibitem[Jasniewicz \& Th\'{e}venin 1994]{jas94} Jasniewicz, G.,
	\& Th\'{e}venin, F. 1994, \aap, 282, 717
\bibitem[Johnson \& Sandage 1955]{joh55} Johnson, H.L., \& Sandage, A.R.
	1955, \apj, 121, 616
\bibitem[Jones et al. 1984]{jon84} Jones, J.E., Alloin, D.M., \&
	Jones, B.J.T. 1984, \apj, 283, 457
\bibitem[J\o rgensen et al. 1992]{jor92} J\o rgensen, U.G., Carlsson, M.,
        \& Johnson, H.R. 1992, \aap, 254, 258
\bibitem[Kim et al. 1998]{kim98} Kim, S., Staveley-Smith, L., Dopita, M.A.,
	Freeman, K.C., Sault, R.J., Kesteven, M.J., \& McConnell, D.
	1998, \apj, 503, 674
\bibitem[Kontizas et al. 1990]{kon90} Kontizas, M., Morgan, D.H.,
	Hatzidimitriou, D., \& Kontizas, E. 1990, \aaps, 84 527
\bibitem[Kraft 1994]{kra94} Kraft, R.P. 1994, \pasp, 106, 553
\bibitem[Kurucz 1999]{kur99} Kurucz, R.L. 1999, HiA, 11, 646; 
	{\it http://cfaku5.harvard.edu/grids.html}
\bibitem[Lee 1977]{lee77} Lee, S.W. 1977, \aaps, 27, 381
\bibitem[Lehnert et al. 1992]{leh92} Lehnert, M.D., Bell, R.A.,
	Hesser, J.E., \& Oke, J.B. 1992, \apj, 395, 466
\bibitem[McWilliam 1997]{mcw97} McWilliam, A. 1997, \araa, 35, 503
\bibitem[Monet 1998]{mon98} Monet, D.G. 1998, \baas, 193, \#120.03
\bibitem[Morell et al. 1992]{mor92} Morell, O., Kallander, D., \&
	Butcher, H.R. 1992, \aap, 259, 543
\bibitem[Nissen et al. 1987]{nis87} Nissen, P.E., Twarog, B.A.,
	\& Crawford, D.L. 1987, \aj, 93, 634
\bibitem[Nikolaev \& Weinberg 2000]{nik00} Nikolaev, S., \& 
	Weinberg, M.D. 2000, \apj, submitted, astro-ph/0003012
\bibitem[Olsen 1999]{ols99} Olsen, K.A.G. 1999, \aj, 117, 2244
\bibitem[Olszewski 1993]{ols93} Olszewski, E.W. 1993, in 
	``The Globular Cluster-Galaxy Connection'', eds.
	G.H. Smith \& J.P. Brodie, (ASP: San Francisco), p. 351
\bibitem[Olszewksi 1999]{edo99} Olszewski, E.W. 1999, in IAU Symp. 190,
	``New Views of the Magellanic Clouds'', eds.
	Y.-H. Chu, N.B. Suntzeff, J.E. Hesser, \& D.A. Bohlender,
	(PASP, San Francisco), p. 292
\bibitem[Olszewski et al. 1991]{ols91} Olszewski, E.W., Schommer, 
	R.A., Suntzeff, N.B., \& Harris, H.C. 1991, \aj, 101, 515
\bibitem[Olszewski et al. 1996]{ols96} Olszewski, E.W., Suntzeff,
	N.B., \& Mateo, M. 1996, \araa, 34, 511
\bibitem[Pagel \& Tautvai\v{s}ien\.{e} 1998]{pag98} Pagel, B.E.J.,
	\& Tautvai\v{s}ien\.{e}, G. 1998, \mnras, 299, 535
\bibitem[Ratnatunga \& Bahcall 1985]{rat85} Ratnatunga, K.U., \&
	Bahcall, J.N. 1985, \apjs, 59, 63
\bibitem[Renzini \& Voli 1981]{ren81} Renzini, A., \& Voli, M. 1981,
	\aap, 94, 175
\bibitem[Richtler 1989]{ric89} Richtler, T. 1989, \aap, 211, 199
\bibitem[Richtler 1990]{ric90} Richtler, T. 1990, \aaps, 86, 103
\bibitem[Richtler et al.\ 1999]{ric99} Richtler, P., Hilker, M.,
	\& Richtler, T. 1999, \aap, 350, 376
\bibitem[Rutledge et al. 1997a]{rut97a} Rutledge, G.A.,
        Hesser, J.E., Stetson, P.B.,
        Mateo, M., Simard, L., Bolte, M., Friel, E., \& Copin, Y.
        1997a, \pasp, 109, 883
\bibitem[Rutledge et al. 1997b]{rut97b} Rutledge, G.A.,
        Hesser, J.E., \& Stetson, P.B. 1997b, \pasp, 109, 907
\bibitem[Salaris et al. 1993]{sal93} Salaris, M., Chieffi, A., 
	\& Straneiro, O. 1993, \apj, 414, 580
\bibitem[Sarajedini 1998]{sar98} Sarajedini, A. 1998, \aj, 116, 738
\bibitem[Sarajedini 1999]{sar99} Sarajedini, A. 1999, \aj, 118, 2321
\bibitem[Schlegel et al. 1998]{sch98} Schlegel, D.J., Finkbeiner,
	D.P., \& Davis, M. 1998, \apj, 500, 525
\bibitem[Smecker-Hane et al. 1999a]{sme99} Smecker-Hane, T.A.,
	Gallagher, J.S., III, Cole, A.A., Tolstoy, E., \&
	Stetson, P.B. 1999a, in IAU Symp. 190, ``New Views of the
	Magellanic Clouds'', eds. Y.-H. Chu, N.B. Suntzeff, 
	J.E. Hesser, \& D.A. Bohlender, (PASP, San Francisco), p. 343
\bibitem[Smecker-Hane et al. 1999b]{sme99a} Smecker-Hane, T.A.,
	Mandushev, G., Hesser, J.E., Stetson, P.B., Da Costa, G.S.,
	\& Hatzidimitriou, D. 1999b, in ``The Spectrophotometric
	Dating of Stars and Galaxies'', eds. I. Hubeny, S. Heap, \&
	R. Cornett, (ASP: San Francisco), p. 159
\bibitem[Smecker-Hane et al. 2000]{sme00} Smecker-Hane, T.A.,
	Cole, A.A., Gallagher, J.S., III, Stetson, P.B., \& 
	Tolstoy, E. 2000, in preparation
\bibitem[Smith 1992]{smi92} Smith, G.H. 1992, \pasp, 104, 894
\bibitem[Stappers et al. 1997]{sta97} Stappers, B.W., Mould, J.R.,
	Sebo, K.M., Holtzman, J.A., Gallagher, J.S., III, Watson, A.M.,
	\& The WFPC2 IDT 1997, \pasp, 109, 292
\bibitem[Stetson 1981]{ste81} Stetson, P.B. 1981, \aj, 86, 1337
\bibitem[Stetson 1987]{ste87} Stetson, P.B. 1987, \pasp, 99, 191
\bibitem[Stetson 1989]{ste89} Stetson, P.B. 1989, HiA, 8, 635
\bibitem[Stetson 1994]{ste94} Stetson, P.B. 1994, \pasp, 106, 250
\bibitem[Stetson \& Harris 1977]{ste77} Stetson, P.B., \& Harris, W.E. 
	1977, \aj, 82, 954
\bibitem[Str\"{o}mgren 1963]{str63} Str\"{o}mgren, B. 1963, in
	``Basic Astronomical Data'', ed. K. Aa. Strand, (U. Chicago
	Pr., Chicago), p. 123
\bibitem[Stryker 1984]{str84} Stryker, L.L. 1984, \apjs, 55, 127
\bibitem[Suntzeff et al. 1992]{sun92} Suntzeff, N.B., Schommer, R.A.,
	Olszewski, E.W., \& Walker, A.R. 1992, \aj, 104, 1743
\bibitem[Suntzeff et al. 1993]{sun93} Suntzeff, N.B.,
        Mateo, M., Terndrup, D.M.,
        Olszewski, E.W., Geisler, D., \& Weller, W. 1993, \apj, 418, 208
\bibitem[Testa et al. 1995]{tes95} Testa, V., Ferraro, F.R., Brocato, E.,
	\& Castellani, V. 1995, \mnras, 275, 474
\bibitem[Thomas 1967]{tho67} Thomas, H.-C. 1967, \zap, 67, 420
\bibitem[Tifft 1964]{tif64} Tifft, W.G. 1964, in IAU-USRI Symp. 20,
	``The Galaxy and the Magellanic Clouds'', eds. F.J. Kerr \&
	A.W. Rodgers, (Austr. Acad. Sci: Canberra), p. 349
\bibitem[Tinsley 1979]{tin79} Tinsley, B.M. 1979, \apj, 229, 1046
\bibitem[Tolstoy \& Saha 1996]{tol96} Tolstoy, E., \& Saha, A. 1996,
	\apj, 462, 672
\bibitem[Vallenari et al. 1996]{val96} Vallenari, A., Chiosi, C., 
	Bertelli, G., Aparicio, A., \& Ortolani, S. 1996, \aap, 309, 767
\bibitem[van den Bergh 1998]{van98} van den Bergh, S. 1998, \apjl, 507, L39
\bibitem[Westerlund 1997]{wes97} Westerlund, B.E. 1997, ``The Magellanic
	Clouds'', (Cambridge Univ. Pr.: Cambridge)
\bibitem[Westerlund et al. 1998]{wes98} Westerlund, B.E., Lundgren, K.,
	Pettersson, B., \& Koziej, E. 1998, \aap, 339, 385
\bibitem[Whitford 1958]{whi58} Whitford, A.E. 1958, \aj, 63, 201
\bibitem[Wyse \& Gilmore 1995]{wys95} Wyse, R.F.G., \& Gilmore,
	G. 1995, \aj, 110, 2771
\bibitem[Zaritsky 1999]{zar99} Zaritsky, D. 1999, \aj, 118, 2824
\bibitem[Zinn \& West 1984]{zin84} Zinn, R., \& West, M.J. 1984,
	\apjs, 55, 45
\end{thebibliography}
\end{document}